\documentstyle[aps,amsfonts,floats]{revtex}
\input psfig.tex 

\def\m180{M_{180}}
\def\ml{M_{\rm lim}}

\def\Rv{R_{\rm vir}}
\def\Rt{R_{\rm ta}}

\def\zt{z_{\rm ta}}

\def\Mv{M_{\rm vir}}
\def\rc{\rho_{\rm crit}}
\def\rcl{\rho_{\rm cluster}}
\def\Oq{\Omega_{\rm Q}}
\def\Om{\Omega_{\rm m}}

\def\Slim{S_{\rm lim}}

\newcommand{\beq}{\begin{equation}}
\newcommand{\eeq}{\end{equation}}
\newcommand{\eea}{\end{eqnarray}}
\newcommand{\bea}{\begin{eqnarray}}

\begin{document}

\title{Constraining cosmological parameters using Sunyaev-Zel'dovich cluster surveys}
\author{Richard A. Battye$^{1}$ and Jochen Weller$^{2}$}

\address{$^{1}$ Jodrell Bank Observatory, University of Manchester, Macclesfield, Cheshire SK11 9DL, U.K. \\ $^{2}$ Institute of Astronomy, Madingley Road, Cambridge CB3 0HA, U.K.}

\date{\today}
\maketitle

\begin{abstract}
We discuss how future cluster surveys can constrain cosmological
parameters with particular reference to the properties of the dark
energy component responsible for the observed
acceleration of the universe by probing the  evolution of the surface
density of clusters as a function of redshift. 
We explain how the abundance of clusters
selected using their Sunyaev-Zel'dovich effect can be computed as a
function of the observed flux and redshift taking into account
observational effects due to a finite beam-size. By constructing 
an idealized set of simulated observations for a fiducial model, we
forecast the likely constraints that might be possible for a variety of
proposed surveys which are assumed to be flux limited. We find that
Sunyaev-Zel'dovich cluster surveys can provide vital complementary information
to those expected from surveys for supernovae.
We analyse the impact of statistical and systematic
uncertainties and find that they only slightly limit
our ability to constrain the equation of state of the dark energy
component. 
\end{abstract}

\section{Introduction}

Sunyaev and Zel'dovich~\cite{Sunyaev:72,Sunyaev:80} (SZ) first noted
that cosmic microwave background (CMB)  photons would be re-scattered
as they pass through the hot intergalactic medium in clusters of
galaxies due to inverse Compton scattering. Since this process
preserves the overall number of photons, one observes a decrement in
CMB temperature in the Rayleigh-Jeans part of the spectrum, and a
corresponding increase at high frequencies above the null frequency of
217 GHz. For a cluster of given mass, the brightness temperature
depends only on the integrated pressure of the gas in the cluster
along the line of sight and not on its distance since the red-shifting
of the photons is exactly balanced by their higher density at the time
of scattering. This not only allows the clusters to be detected at
much higher redshifts than using standard optical and X-ray
measurements, it also makes any detection much less susceptible to the
internal dynamics of the cluster, providing a reliable method for
detecting clusters in a blank field survey. 

To date the SZ effect has been mapped in a small number of clusters
with targeted observations using either a single
dish~\cite{Birkinshaw:86,Birkinshaw:91,Hughes:98}, or an
interferometer~\cite{Jones:93,Grainge:93,Carlstrom:96,Grainge:96,Carlstrom:98,Reese:00,Grego:00,Patel:00,Joy:01,Grego:01}.  
In the very near future blank field surveys will be performed on
dedicated instruments with arcminute resolution which should yield
flux limited samples of clusters at a variety of frequencies and
resolutions over large areas (see, for
example,~\cite{Bolocam:1,Bolocam:2,AMI:1,AMI:2,SZA:1,SZA:2,SZA:3,Amiba,OCRA,SPT:1,SPT:2,ACT,APEX}
and references therein). Moreover, instruments originally designed to
make small scale CMB anisotropy measurements such as the Very Small
Array (VSA)~\cite{VSA}, Cosmic Background Imager (CBI)~\cite{CBI} and
those on board the Planck Surveyor~\cite{Aghanim:97,Kay:01} will also
inadvertently provide information, albeit less efficiently due to
their large beam sizes~\cite{BWproc}, though the Planck Surveyor can
balance this disadvantage with its full sky coverage and the range of frequencies available.

The limiting mass of a survey as a function of redshift can be
computed from the limiting flux assuming the cluster is virialized
and, hence, that there exists a relationship between the mass of the
cluster and the temperature of the gas within it. Therefore, the survey
yield is a calculable function of the cosmological parameters and the
process of halo formation. If we believe the physics of halo formation
is understood, which in the case of cluster size halos should just
involve gravitational physics, then the number of clusters as a
function of redshift can be used to constrain cosmological
parameters~\cite{Thomas:89,Oukbir:92,Scaramella:93,Barbosa:96,Viana:96,Eke:96,Wang:98,Pierpaoli:01,Haiman:01a,Holder:01,Benson:02,Weller:02b}.
More precisely the number of objects  will depend on the comoving
volume element, the angular diameter distance and the rate of growth
of perturbations, all of  which depend sensitively on the late-time
evolution of the cosmological scale factor $a(t)$. 

The study of the late-time evolution of the universe has become an
important part of many observational programs ever since
observations of type Ia supernovae (SNe) suggested that the expansion
of the universe might be
accelerating~\cite{Perlmutter:97,Riess:98,Perlmutter:99a,Riess:01}. It
was found that a combined initial sample of 42 SNe with $\langle
z\rangle\approx 0.6$ were much dimmer at high redshift than required
by a matter dominated Einstein-de Sitter universe, with the simplest
remedy being the inclusion of Einstein's `greatest blunder', the
cosmological constant $\Lambda$ which gives rise to acceleration by
virtue of the negative pressure associated with vacuum energy. Subsequent observations of more SNe have only strengthened this conclusion. These
observations probe the magnitude-redshift relation by assuming the SNe
are standard candles.

However, this is notoriously difficult to incorporate into a realistic
particle physics theory since estimates of the vacuum energy, while
very rarely zero, are substantially larger than the observed
value. This lead to the proposal of a dynamical dark energy component
with negative pressure, often called `Quintessence', due to a scalar
field~\cite{Wetterich:88,Ratra:88,Peebles:88,Wetterich:95,Frieman:95,Coble:97,Ferreira:97,Copeland:98,Caldwell:98,Zlatev:99,Steinhardt:99,Brax:99,AS:00}. Although this idea is not without its own problems, it provides an interesting
theoretical framework to parameterize and test the idea of dark energy.

If one takes Quintessence seriously one has to accept the fact that
the equation of state parameter $w_{\rm Q}=p_{\rm Q}/\rho_{\rm Q}$ is not
only negative, but that it can evolve with time. There are many specific dark
energy models, none of which are theoretically compelling and,
therefore, in order to test them one requires some kind of parameterization of  this evolution. There has been some debate in the literature about what is the best way to
do this: suffice to say that it is difficult to imagine one which
accurately represents every model and it is often best to tailor the
parameterization to the type of observations under
consideration~\cite{Efstathiou:99,Huterer:99,Saini:99,Maor:01,Astier:00,Weller:01,Weller:01a,Sahni:02,Huterer:02}. Since the clusters under discussion here will have redshift of less than 1.5, a Taylor expansion of the form
$w_{{\rm Q}}=w_0+w_1z+..\,,$ where $w_0$ and $w_1$ are constants is a
sensible form to consider. 

An important feature of SZ surveys, and similar observations that
probe number counts, is that their dependence on the expansion rate of
the universe is very different to that of SNe observations, that is, they
are complimentary\footnote{Note that measurements of the angular
diameter distance relation at low redshifts using, for example,
gravitational lensing statistics~\cite{class1,class2} have exactly the
same degeneracy as those using the magnitude-redshift relation. These
can be used as a direct test of the veracity of SNe surveys.} and,
hence, can be used to break parameter degeneracies. A degeneracy
between $w_0$ and $w_1$ is present for both SZ and SNe
surveys. However, the shape of the likelihood contours in the
$w_0$-$w_1$ plane is very different~\cite{Weller:02b}, allowing one to
use the two together to make more substantial statements as to the
nature of the dark energy. 

In this paper we will discuss using the evolution of the surface
density as a a function of redshift. A
particular important ingredient for using number counts of clusters to
constrain cosmology is the ability to measure their redshift. Ideally
spectroscopic redshifts ($\Delta z \approx 0.01$) would be available
for each cluster. However, it may only be possible to get photometric
redshift ($\Delta z \approx 0.1)$. Moreover, at present it is
difficult to estimate redshifts for objects with $z\sim 1$. The Sloan
Digital Sky Survey (SDSS)~\cite{SDSS:1,SDSS:2} and the Visible and
Infrared Survey Telescope for Astronomy (VISTA)~\cite{VISTA} will
measure the redshift of many hundreds of thousands of galaxies and
will provide this information at least to some degree. Except where
explicitly stated we will use $\Delta z=0.1$ and $z_{\rm max}=1.5$,
optimistically assuming techniques to find redshifts for very distant
objects can be developed on a similar timescale to instruments under
discussion here. 

In section \ref{sec:redshift} we will discuss the principal cosmology
dependence of the surface density of clusters on cosmology, before
going on to discuss the relation between the limiting mass and the
survey set up in section~\ref{sec:mass}. We will then perform a mock
likelihood analysis for six proposed survey setups in
section~\ref{sec:likelihood} and discuss statistical and systematic uncertainties in our assumptions in section~\ref{sec:systematics}. Note that
throughout this paper algebraic relations will be expressed in
natural units where $c=\hbar=k_{\rm B}=1$. 

\section{Surface density of clusters}\label{sec:redshift}

There exists a plethora of dark
energy models none of which are particularly compelling. We will only  discuss models which are minimally coupled to gravity.
Since, a priori, there is no theoretical preference
for any particular model we take a phenomenological approach and
characterize the dark energy component by a linear evolving equation
of state~\cite{Maor:01,Astier:00,Weller:01}
\beq
	w_{\rm Q}=w_0+w_1z \, .
\label{eqn:weqn}
\eeq
This is sufficient since we are only interested in the low redshift
behaviour of the dark energy component and furthermore
(\ref{eqn:weqn}) approximates a wide range of dark energy models
adequately at low redshifts~\cite{Maor:01,Astier:00,Weller:01,Weller:01a}.

In order to predict the surface density of clusters of a mass limited
Sunyaev-Zel'dovich survey we need to compute 
\beq
	\frac{dN}{dz} =
{\Delta\Omega}\frac{dV}{dzd\Omega}(z)\int\limits_{M_{\rm 
lim}(z)}^\infty \frac{dn}{dM}\,dM\, ,
\eeq
where $dV/(dzd\Omega)= [r(z)]^2/H$ is the comoving volume in a flat
universe, with $r(z)=\int_0^z H^{-1}(z^{\prime})dz^{\prime}$ the
coordinate distance, 
$\Delta\Omega$ is the angular sky coverage of the survey and $dn/dM$
is the comoving number density of objects with mass $M$ and redshift $z$, sometimes called the mass function. $\ml$ is the limiting mass, which will depend in general on the parameters of the survey. In this section, we will assume that $\ml$ is constant and does not depend on the cosmological
parameters. In the next section, however, we will drop this assumption
and model $\ml$ in a more realistic way for our mock likelihood analysis.

We will use the comoving number density calibrated using a series of N-body
simulations performed by the VIRGO consortium~\cite{Evrard:02}, with
\beq
\frac{dn}{dM}\left(z,M\right) = - 0.22
\frac{\rho_m(t_0)}{M\sigma_{M}}\frac{d\sigma_M}{dM}
\exp\left\{-|0.73-\log\left[D(z)\sigma_M\right]|^{3.86}\right\}\, ,
\label{eqn:conum}
\eeq
where the mass, $M$ of the object is defined to be that inside
a spherical over-density of $\Delta=200$ times the critical background
density, that is, $M_{200}=4\pi R_{200}^3 200 \rc(z)/3$.
$D(z)$ is the growth factor normalized to have $D(0)=1$. We obtain the growth factor by
solving numerically the perturbation equation for the matter fluctuations
\beq
	\delta_m^{\prime\prime}
+\frac{3}{2}a^{-1}\left[1-w(a)\left(1-\Omega_{\rm
m}(a)\right)\right]\delta_m^\prime-\frac{3}{2}a^{-2}\Omega_{\rm
m}(a)\delta_m =0\, ,
\eeq
where the prime denotes the derivative with respect to the scale
factor $a$ and we choose our initial conditions to be setup in the matter
dominated era with $\delta_m(a_i) =a_i$ and $\delta_m^\prime(a_i) =
1$, where $z_i=30$ is sufficient. The growth factor is then given by $D(z)
=\delta_m(z)/\delta_m(z=0)$. Furthermore, $\rho_m(t_0)$ is the present day matter density, $\Om(a)=\rho_m(a)/\rc(a)$ is the matter density in
units of the critical density and  $\sigma_M$ is the over-density with mass $M$ today, 
where
\beq
\sigma_R^2 = \int_0^\infty W^2(kR) \Delta^2(k) \frac{dk}{k}
\label{eqn:sig}
\eeq
and $M=4\pi\rho_m(t_0)R^3/3$. The window function in (\ref{eqn:sig}) is that of a spherical top hat with $W(x) =
3(\sin(x)/x^3-\cos(x)/x^2)$.  

The final missing ingredient is the linear power spectrum $P(k)$, where $\Delta^2(k) = 4\pi k^3P(k)$. We define the transfer function $T(k,z)$ via $P(k) = A
k^n[T(k,z)]^2$, where $n$ is the
spectral index. CMBFAST~\cite{Seljak:96} is used to compute the shape of the
transfer functions in $k$ space, where it is sufficient to approximate the
perturbations of a dark energy model with the corresponding 
$\Lambda$CDM model, since
clustering on small scales is {\em not} affected by the presence of a
dark energy component. We fix the baryonic component  to have 
$\Omega_{\rm b}h^2 = 0.019$ in agreement with Big Bang Nucleosynthesis
constraints~\cite{Copi:95,Burles:98}. To normalize the power spectrum, 
we use $\sigma_8$, the over-density in a spherical
region of $8h^{-1} {\rm Mpc}$, which is traditionally used to quantify the amplitude of the power spectrum on small scales. 

For our fiducial model we choose
$\sigma_8=0.9$ as suggested by recent CMB measurements~\cite{wmap}.
Note that this value is higher than the one inferred from recent X-ray
measurements~\cite{Seljak:01,Viana:02,Allen:02b}. One might think
that this discrepancy is not very worrying since $\sigma_8$ is a
parameter which we are trying to determine. However, since the number
of objects found in a given survey is very sensitive to $\sigma_8$ and
the statistics is Poisson distributed, using a much smaller value,
$\sigma_8=0.7$ say, would substantially weaken our conclusions. Also
we assume that the universe is flat and $H_0   
= 72\,{\rm km}\,{\rm sec}^{-1}{\rm Mpc}^{-1}$~\cite{Freedman:01},
$\Omega_m(t_0)=0.3$ and a spectral index
$n=1$~\cite{Bean:02,Lewis:02,Allen:02a}. For the 
dark energy equation of state parameters we choose $w_0=-0.8$
and $w_1=0.3$ which is allowed by current data
~\cite{Garnavich:98,Perlmutter:99b,Bean:02,Lewis:02} and is
compatible with  an interesting  Supergravity related Quintessence
model~\cite{Brax:99,Weller:01a}. 

\begin{figure}[!h] 
\centerline{\psfig{file=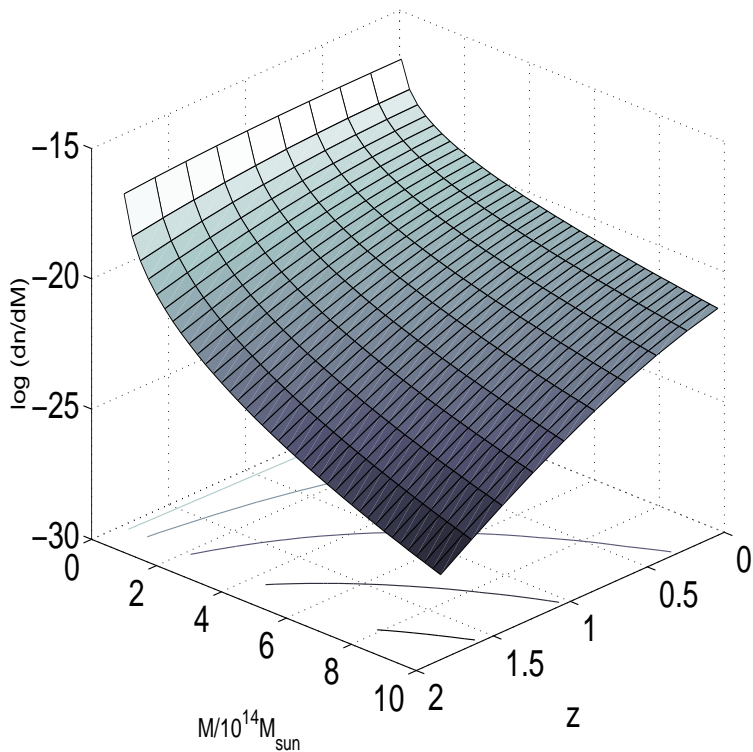,width=10cm,height=10cm}}
\caption{The comoving mass density $dn/dM$ in units of $[{\rm
Mpc}^3M_\odot]^{-1}$ from simulations of the VIRGO
consortium~\protect\cite{Evrard:02} using a $\Lambda$CDM
cosmology. The lines on the base of the plot represent contours of
constant $dn/dM$.} 
\label{fig:dndM}
\end{figure}

In Fig.~\ref{fig:dndM} we show the comoving mass density for a
$\Lambda$CDM cosmology (w=-1). The rapid decrease for large cluster masses
and  
large redshifts is clearly visible and we expect a low number of
clusters in these ranges. This choice of comoving number
density is  
similar to the usual Press-Schechter (PS)
formalism~\cite{Press:74}. However, the PS approach shows 
some significant differences from the observed and simulated comoving
number density. The simulations show less low  and an increased
number of high mass clusters (see also Fig.~\ref{fig:dNdz_mass})
compared to the PS prescription. A much
better fit is achieved from using PS if one assumes an 
aspherical collapse~\cite{Sheth:01,Jenkins:01,Pierpaoli:01}. Also note
that the mass function has an
uncertainty~\cite{Pierpaoli:01,Jenkins:01,White:01,Evrard:02,White:02}
and we will come to this problem when we discuss systematic errors in
section~\ref{sec:systematics}. 

\begin{figure}[!h]
\setlength{\unitlength}{1cm}
\centerline{\psfig{file=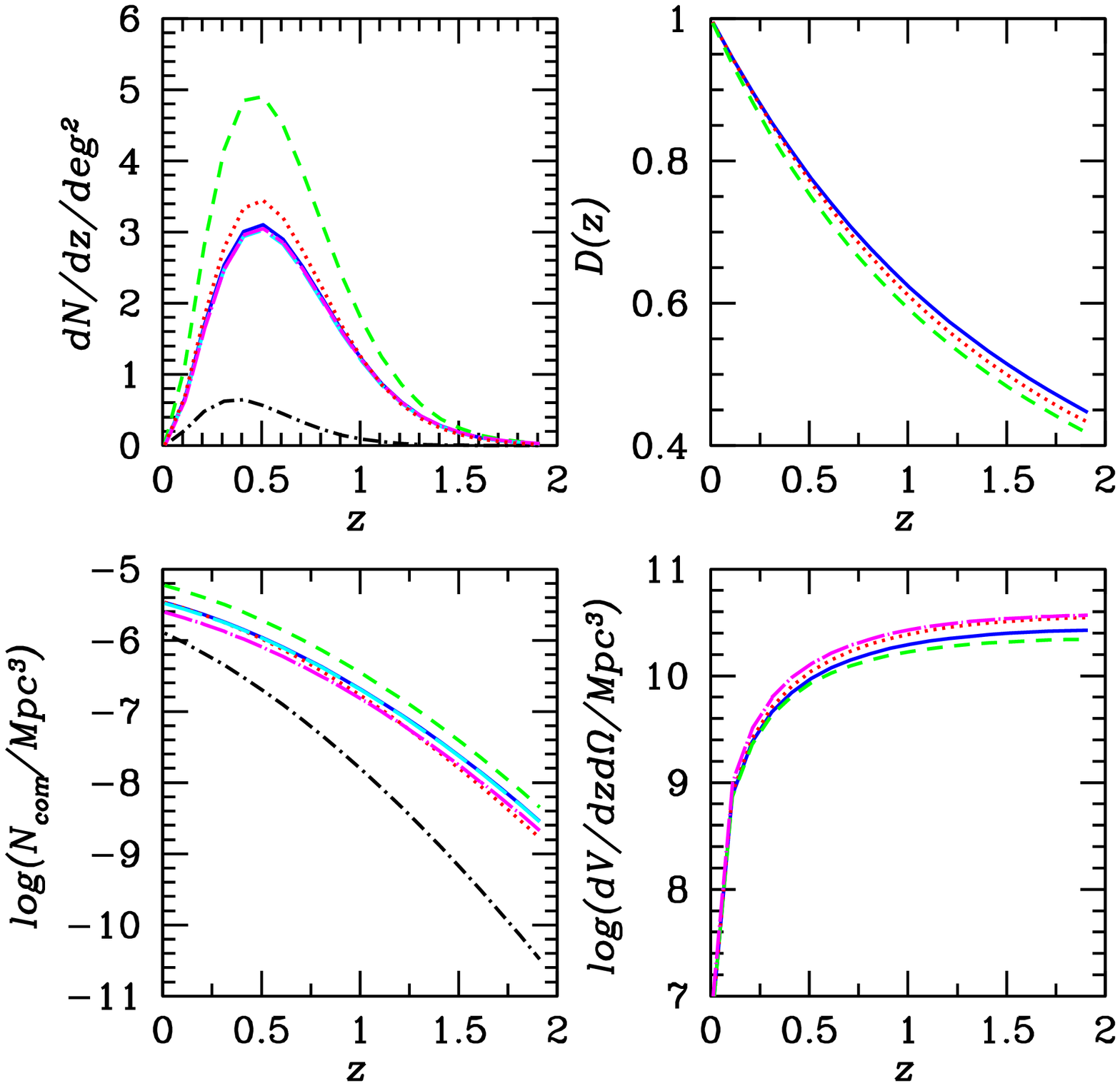,width=12cm,height=10cm}}
\caption{In the top left we show the surface density of clusters above
a mass limit of $M_{\rm lim} =2.35\times 10^{14} M_\odot$ 
as a function of redshift. The top right is the normalized growth
factor $D(z)$, the lower left the comoving number of clusters $N_{\rm
com}(z)$ and the lower right is the volume element
$dV/(dzd\Omega$). The solid line is the fiducial cosmology (see text), 
the dotted line a $\Lambda$CDM cosmology, the short dashed line is for
$\Om(t_0)=0.4$, the long dashed line for $n=0.9$ (almost underneath the
solid line), the dot - short dash line for $\sigma_8 = 0.72$ and the dot - long dash line for $H_0=65\,{\rm km}\,{\rm sec}^{-1}{\rm Mpc}^{-1}$.}
\label{fig:dncosmo}
\end{figure}

In the following we will discuss the dependence on cosmology of the
number of objects per square degree above a given mass limit $M_{\rm
lim}= 2.35\times 10^{14} M_\odot$. We will see in
section~\ref{sec:mass} that $M_{\rm lim}=M_{\rm lim}(z)$, but using a
fixed limiting mass should provide some intuitive information. In
Fig.~\ref{fig:dncosmo} we present the 
surface density of clusters (top left) and its constituents, the
comoving number density, $N_{\rm com} = \int_{M_{\rm 
lim}(z)}^\infty \frac{dn}{dM}\,dM $, (lower left), the growth factor
(top right) and the volume (lower right).
As a base for comparison we use the
fiducial cosmology described above and then change subsequently one
parameter.  As expected the strongest dependence is on $\sigma_8$ and
$\Omega_m(t_0)$. However, we see a weak, but nonetheless significant
dependence on the equation of state parameters $w_0$ and $w_1$. The
dependence on the spectral index $n$ is very weak and hence we fix
$n=1.0$ in the subsequent analysis. We also see that the strongest
dependence must come from the linear growth factor (upper right
panel), since minute changes lead to significant changes in the surface
density. This is most obvious from the change in the $\Om$ (dashed
line), where the volume element (lower right panel) would predict a
lower number of clusters, which is balanced by the faster growing
linear perturbations (top right). This is clear, since the dependence
of the surface density on the growth factor is exponential (Eqn.~\ref{eqn:conum}).

\section{The limiting mass}\label{sec:mass}

In the previous section we assumed that the mass limit of the survey
is constant. However, in a realistic observing situation the survey is
limited by the flux and not mass. Hence, we need a relation between
the flux limit $\Slim$ and the mass limit $M_{\rm lim}$. We should
note that, contrary to what is often explicitly stated or implied in
the literature, the surveys are not mass limited in the sense that
there is single limiting mass which applies across a wide range of
redshifts. If one were to use a fixed limiting mass, one would be
either forced to throw out many objects in order to make the survey
complete to some high threshold, or put up with lack of completeness
at both low and high redshifts with a smaller threshold. Here, we will
discuss how to compute the limiting mass by first making the extreme,
and generally incorrect assumption, that the cluster is point-like in the telescope beam. We will then extend our results to
clusters which are resolved. A more detailed discussion of this
problem can be found in the relevant section of the companion
paper~\cite{Battye:02} 

\subsection{Principal dependence on cosmology}
We first construct the relation between the virial mass and flux limit of the 
survey assuming that the source is point-like  in the telescope beam. The
total flux density and the brightness
temperature are related by~\cite{Sunyaev:72,Sunyaev:80,Birkinshaw:99} 
\beq
S_{\nu}=2\nu^2 \Delta T f(x) =2\nu^2 T_{\rm CMB}f(x)Y\,,
\label{eqn:flux}
\eeq
with $f(x) = x^2e^xg(x)/(e^x-1)^2$ and $g(x) = x/{\rm tanh}(x/2)-4$,
where $x = 2\pi\nu/T_{\rm CMB}$ is the dimensionless frequency. The $Y$
parameter is given by the integrated y-distortion $Y=\int d\Omega
y(\theta)$, with
\beq
	y(\theta) ={\sigma_{\rm T}\over m_{\rm e}}\int dl\,n_{\rm
e}\,T_{\rm e}={\langle T_{\rm e}\rangle_{n}\over m_{\rm e}}\tau_{\rm e}(\theta)\, ,
\label{eqn:ydist}
\eeq
the line of sight integral over the gas pressure in the
cluster. The Thomson scattering cross section is $\sigma_{\rm T}$ and
$m_{\rm e}$ is the electron mass. We have introduced the electron density
weighted average temperature $\left<T_{\rm e}\right>_{n}=\int dl\, n_{\rm
e}T_{\rm e}/\int dl\, n_{\rm e}$, and $\tau_{\rm e}(\theta) =
\sigma_{\rm T} \int dl\, n_{\rm e}$ the optical depth.

It is conventional to normalize the optical depth in terms of the virial mass of the cluster using 
\begin{equation}
{M_{\rm vir}f_{\rm gas}\over \mu_{\rm e}m_{\rm p}}=\int d^3r\, n_{\rm
e}({\bf r})=d_A^2\int d\Omega dl n_{\rm e}\,,
\end{equation}
from which one can deduce that 
\begin{equation}
\tau_{\rm e}(0)=\sigma_{\rm T}{1\over d_{\rm A}^2}{f_{\rm gas} M_{\rm vir}\over \mu_{\rm e}m_{\rm p}}\left(\int \zeta(\theta)d\Omega\right)^{-1}\,,
\end{equation}
where $\tau_{\rm e}(\theta)=\tau_{\rm e}(0)\zeta(\theta)$, $f_{\rm
gas}$ is the intra cluster gas fraction, which we choose $f_{\rm
gas}=0.12$ throughout this paper~\cite{Mohr:99}, $\mu_{\rm e} = 1.143$ the
mean mass per electron and $m_{\rm p}$ the proton mass.
In the general the profile $\zeta(\theta)$ will depend on the virial radius $R_{\rm vir}$
and redshift $z$ via the angular diameter distance, which for a flat universe is given by 
$d_A(z)=(1+z)^{-1}\int_0^zH^{-1}(z^\prime)\,dz^{\prime}$. Using this normalization, it is easy to see from this that the flux density is given by  
\begin{equation}
S_{\nu}={2\nu^2T_{\rm CMB}\langle T_{\rm e}\rangle_{n}\over m_{\rm e}}{f_{\rm gas}M_{\rm vir}\over \mu_{\rm e}m_{\rm p}} {1\over d_{\rm A}^2}\sigma_{\rm T}f(x)\,,
\label{eqn:finalflux}
\end{equation}

In order to solve (\ref{eqn:finalflux}) for the mass for a given flux
limit, we need to know the relation between mass $\Mv$ and the cluster
temperature $\langle T_{\rm e}\rangle_{n}$. We assume that
clusters are virialized objects which are in thermal equilibrium. This
assumption seems to be confirmed by recent X-ray observations~\cite{Allen:02b}, though it is expected that due to ongoing mergers and
heat input, this assumption does not hold for high redshifts~\cite{Verde:01}; an issue we will discuss in section
\ref{sec:systematics}. If we assume that the
cluster is virialized the kinetic energy at virialization $E^{\rm
kin}_{\rm vir}$ is given by
\beq
	E^{\rm kin}_{\rm vir} = -\frac{1}{2}U_{G} + U_{{\rm Q}}\, ,
\eeq
with $U_G = -3G\Mv^2/(5\Rv)$ the potential energy due to the gravity
of the matter. By integrating $\epsilon_{\rm Q}=\rho_{\rm Q}+3p_{\rm Q}$ over the sphere, with equation of state factor $w$ and assuming that $\rho_{{\rm Q}}$ is smooth on cluster scales, we obtain the general expression for the potential energy 
\beq
	U_Q = \frac{1+3w}{10}4\pi G \rho_{{\rm Q}}(a) \Rv^2\Mv \, .
\label{eqn:potQ}
\eeq
Note that (\ref{eqn:potQ}) is the generalization of the well known result for
a cosmological constant~\cite{Lahav:91}, and is different to the
form used in~\cite{Wang:98}. We can relate the virial mass of the
cluster to the virial radius, by assuming spherical symmetry $M_{\rm
vir} = 4\pi\Rv^3\rcl/3$, where $\rcl$ is the mean density of the
cluster. The mean kinetic energy $E^{\rm kin}_{\rm
vir}=\Mv\left<V_{\rm vir}^2\right>/2$ is given in terms of the root
mean square of the velocity dispersion of the cluster. If the cluster
is in thermal equilibrium, the virialization temperature is given by~\cite{Lilje:92,Bryan:98,Afshordi:01} 
\beq
\langle T_{\rm e}\rangle_{n} = \frac{1}{3}\mu
m_{\rm p} \langle V_{\rm vir}^2 \rangle \, ,
\eeq
with $\mu=0.59$ the mean mass per particle. Hence, we deduce that  
\beq
	\frac{\Mv}{10^{15}h^{-1}M_\odot} =
\left(\frac{\left<T_{\rm e}\right>_n}{T_*}\right)^{3/2} \left(\Delta_c
E(z)^2\right)^{-1/2}\left[1+\left(1+3w\right)\frac{\Oq(z)}{\Delta_c}\right]^{-3/2}\,
,
\label{eqn:mt}
\eeq
where $T_*$ is a normalization factor which can be deduced to be $T_*\approx 0.5$ under the
assumption that the process of virialization only involves
gravitational heating . Both simulations~\cite{Bryan:98,Pierpaoli:01}
and observations appear to be somewhat inconsistent with this value.
For our subsequent analysis we choose 
$T_*=1.6$ and we will come back to this in section
\ref{sec:mtsys}. Note that we have expressed the mean
cluster density in 
terms of the over-density $\Delta_c$ of the cluster at virialization, with
$\rcl=\Delta_c\rc$.

We can solve numerically for the over-density at virialization
by applying the spherical collapse model
~\cite{Patridge:67,Gunn:72,Viana:96,Wang:98}. We follow essentially the
discussion in \cite{Wang:98} where the background cosmology is
described by the Friedman equation
\beq
	\left(\frac{{\dot a}}{a}\right)^2 =
\frac{8\pi G}{3}\left(\rho_m+\rho_Q\right)\, .
\label{eqn:Fried}
\eeq
and the collapsing cluster of radius $R$ by
\beq
\frac{{\ddot{R}}}{R} = - {4\pi G\over 3}\left[\left(1+3w\right)\rho_Q +\rcl\right]\,.
\label{eqn:Rc}
\eeq
As before, we  make the standard assumption
that the dark energy component does not clump on the
relevant scales. We can solve this coupled system numerically and
obtain the density of the cluster at turnaround $t=t_{\rm ta}$, $\rcl=\zeta_{\rm ta} \rho_{\rm m}$, and obtain a best fit for the over-density
\beq
	\zeta_{\rm ta}(\zt) \approx
\left(\frac{3\pi}{4}\right)^2\Omega_m(\zt)^{-0.79+0.26\Omega_m(\zt)-0.08(w_0-w_1)-0.21w_1(1+\zt)}\, .
\eeq
We can then scale this result to the time of
virialization, $t=t_{\rm v} = 2t_{\rm
ta}$, with
\beq
	\Delta_c(z_{\rm v}) = \zeta_{\rm ta}(\zt)\Omega_{\rm m}(z)
\left(\frac{\Rt}{\Rv}\right)^3\left(\frac{1+\zt}{1+z_{\rm v}}\right)^3\, ,
\eeq
where we have introduced
\beq
	\frac{\Rv}{\Rt} \approx \frac{1-\eta_{\rm v}/2}{2+\eta_{\rm t}-3/2\eta_{\rm v}}\, ,
\eeq
the ratio of the radius at virialization (or collapse) of the cluster
to the radius at turn around with
\bea
	\eta_{\rm t} & = &  -\left[1+3w(\zt)\right]
\frac{\Oq(\zt)}{\zeta_{\rm ta}\Omega_m(\zt)}\,,\\
	\eta_{\rm v} &=& -\left[1+3w(z_{\rm v})\right] \frac{\Oq(z_{\rm v})}{\zeta_{\rm ta}\Omega_m(z_{\rm v})}\left(\frac{1+z_{\rm v}}{1+\zt}\right)^3
\, .
\eea
Note that again these expressions differ from~\cite{Wang:98} because
of the generalized potential for the dark energy component in
(\ref{eqn:potQ}). We have now all the necessary ingredients to
calculate the virial mass of a cluster for a given Sunyaev-Zel'dovich
flux decrement. However in order to calculate the limiting mass in
(\ref{eqn:conum}) we need to transform the virial mass from an
over-density of $\Delta_c$ to an over density of $\Delta = 200$. In order to do this we assume that the matter in the cluster is
distributed according to a Navarro-Frenk-White (NFW) profile
\cite{Navarro:97}, with a concentration parameter $c=5$. This allows
us to rescale the cluster mass from $\Mv$ to $M_{200}$. 
\begin{figure}[!h] 
\setlength{\unitlength}{1cm}
\centerline{\psfig{file=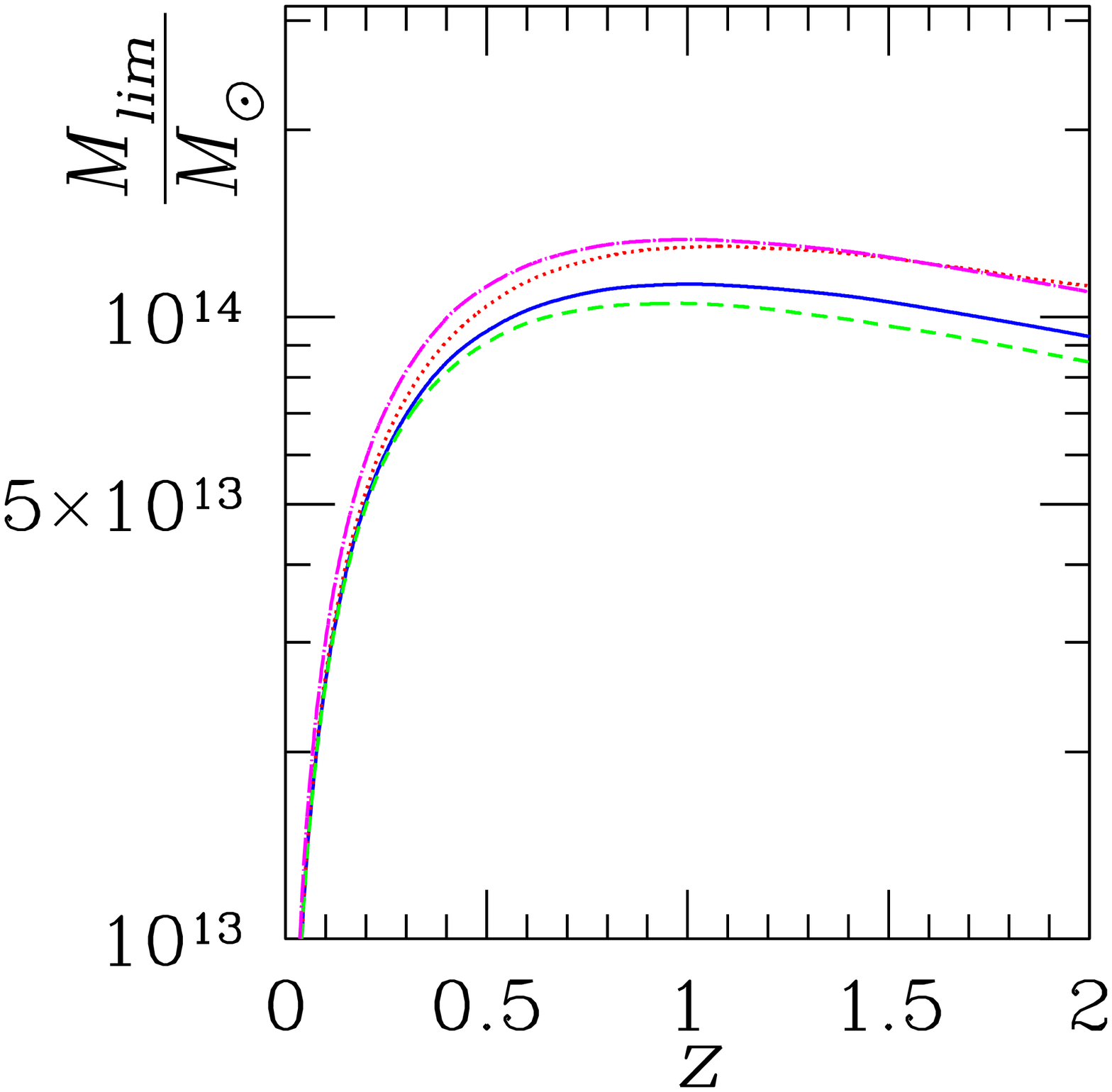,width=7cm,height=7cm}\psfig{file=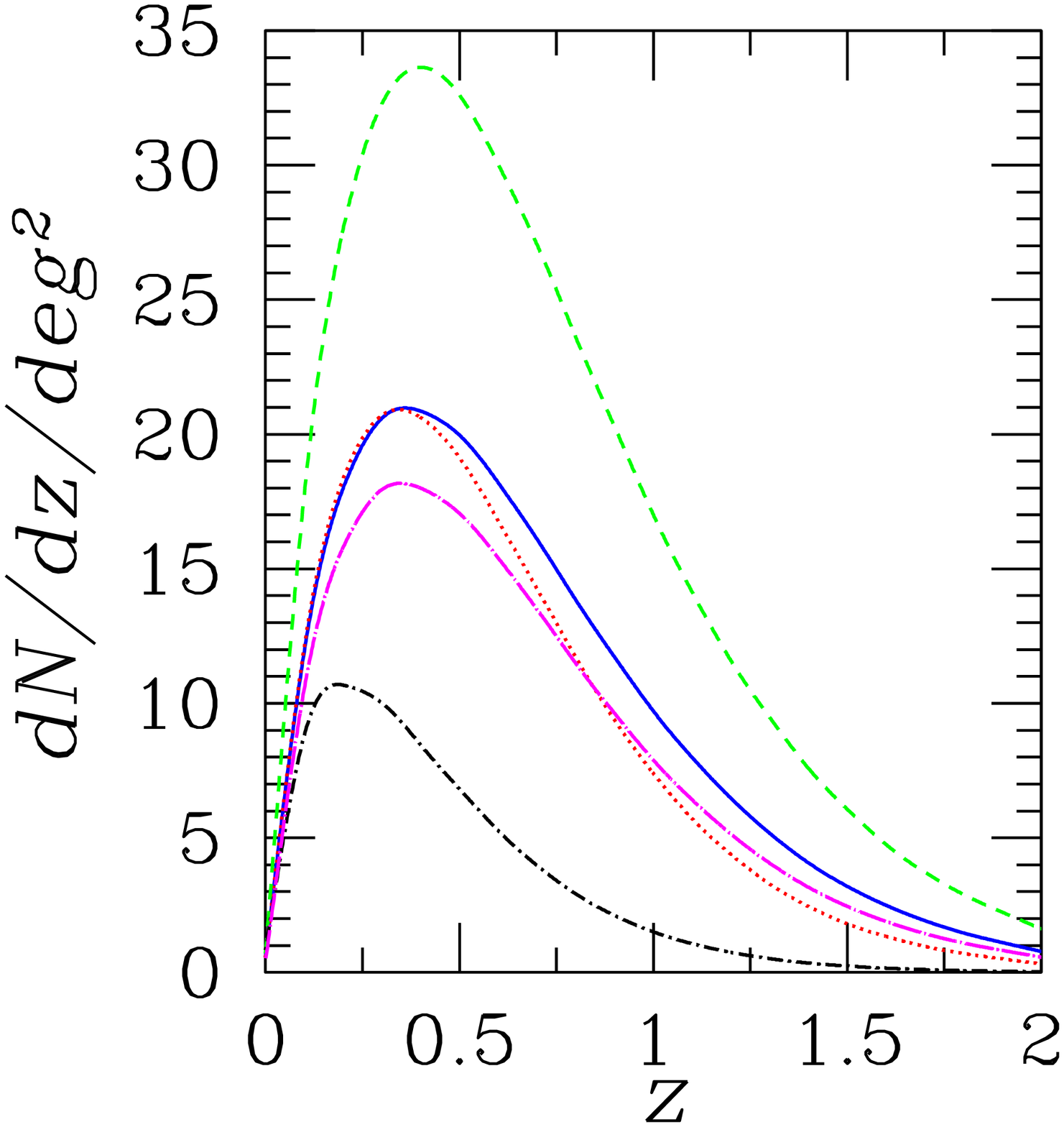,width=8cm,height=7cm}}
\caption{The limiting mass (left) and surface density of
clusters (right) in the point source approximation for a survey with
limiting flux density $S_{\rm lim}=4.4\,{\rm mJy}$ and frequency
$\nu=150\, {\rm GHz}$. The different 
lines correspond to different cosmological parameters with the key the
same as in Fig.~\ref{fig:dncosmo}.}
\label{fig:dnpoint}
\end{figure}
In Fig.~\ref{fig:dnpoint} we illustrate the limiting mass and redshift
distribution of clusters using the point source approximation for the
same cosmologies as in Fig.~\ref{fig:dncosmo}. First, notice that
the limiting mass depends on the cosmological parameters. As well as
the standard dependence on $H_0$, we see dependence on $\Om$ and $w_0$
which would be ignored by assuming that the limiting mass is
constant. Note, that the dependence of $M_{\rm lim}$ on $w_0$ now
lowers the overall number of clusters (dotted line), balancing the slight increase
of the number of clusters for the 
$\Lambda$CDM universe, compared to the fiducial model in the case of a
fixed limiting mass (see Fig.~\ref{fig:dncosmo}). For an increasing
$\Om$ the limiting mass is decreasing resulting in an even more
enhanced number of clusters compared to the fiducial model. Moreover, in
this approximation the limiting mass goes to zero at low
redshifts. Although this will be seen to be unphysical when telescope
beams are small, it will be a good approximation when the beam is
large. The dependence of the limiting mass on the cosmological
parameters feeds into the predictions for $dN/dz$, but the effects of
parameters such as $\sigma_8$ and $n$ is unchanged from that for a
constant limiting mass. 

\subsection{Extended clusters and survey beam}
So far we have not accounted for the possibility of clusters being extended
objects within the telescope beam which could lead to a loss of flux. If there is a beam $B(\theta)$ present, integrated y-distortion in
(\ref{eqn:flux}) becomes~\cite{Bartlett:00} 
\beq
	Y=\int d\Omega\, y(\theta) \quad \to \quad Y = \frac{\int d\Omega\,
B(\theta)y(\theta)}{\int d\Omega B(\theta)}\, ,
\eeq
where we assume a Gaussian profile of the beam, with
$B(\theta)=\exp[-\theta^2/(2\sigma_b^2)]$ and  $\sigma_b$ is related to the full-width-half-maximum (FWHM) of the beam by $\sigma_b =
\theta_{\rm FWHM}/\sqrt{8\ln 2}$. The equation for the observed flux for a beam directly over the centre of the cluster is then~\cite{Battye:02}
\beq
S_{\nu}={2\nu^2T_{\rm CMB}\langle T_{\rm e}\rangle_{n}\over
m_{\rm e}}{f_{\rm gas}M_{\rm vir}\over \mu_{\rm e}m_{\rm p}}
{1\over d_{\rm A}^2}\sigma_{\rm T}f(x){\cal I}(B,\zeta)\,,
\label{eqn:fluxbeam}
\eeq
with
\beq
{\cal I}(B,\zeta)={\int d\Omega B(\theta)\zeta(\theta)\over
\int d\Omega B(\theta) \int d\Omega \zeta(\theta)}\,.
\eeq
In general this function will depend on $\Mv$ via $\Rv$, so a simple
analytical solution for $M_{\rm vir}$ in terms of $S_{\rm lim}$ and $z$ 
is not possible and therefore we must solve
(\ref{eqn:fluxbeam}) numerically using the Newton-Raphson method. The expression (\ref{eqn:fluxbeam}) is the
flux in a single beam, however, if the cluster is larger than one beam area
, beams can be combined with a resultant increase in the noise; a process often known as beam smoothing. The optimum number of beams is given by 
\beq
	N_{\rm b} = \frac{\int d\Omega\,\zeta(\theta)}{\int
d\Omega\, B(\theta)}\,,
\label{eqn:npix}
\eeq
as discussed in \cite{Battye:02}. In reality $N_{\rm b}$ must be an integer, but we will allow it here to take any value greater than one. It was shown in \cite{Battye:02} that this does not unduly exaggerate the predicted number of clusters detected.  

In most situations there will be a limit to the extent to which one can beam smooth, for example, due to the primary CMB anisotropy. In the subsequent discussion we will  assume that $\theta_{\rm FWHM}^{\rm max}=20^{\prime}$. Note that for
multi-frequency observations like Planck it may be   possible to
effectively filter out the SZ signal due its distinct
spectral signature. This in principle would allow to combine 
even more pixels, however we do not include this effect in our
analysis and we do not expect it to be significant, since it 
will be important only for objects at low redshifts. 

We assume that the cluster is in hydrostatic equilibrium and can be
described by a truncated isothermal
$\beta$-model~\cite{Cavaliere:76,Molnar:00}, with an 
electron density distribution of 
\beq
n_{\rm e}(r)=\left\{
\begin{array}{c@{\qquad}l}
	n_0\left(1+\frac{r^2}{R_{\rm c}^2}\right)^{-\frac{3\beta}{2}} &
	r<\Rv \\
        0 & r\ge\Rv
\end{array}\, ,
\right.
\eeq
where $R_{\rm c}$ is the core radius of the cluster defined in terms of the
virial radius $R_{\rm c} = \Rv/\alpha$. The projected profile function is
then
\beq
	\zeta(\theta) =
\left(1+\frac{\theta^2}{\theta_c^2}\right)^{\frac{1}{2}-\frac{3}{2}\beta}\,\frac{J\left[\left(\frac{\alpha^2-\theta^2/\theta_c^2}{1+\theta^2/\theta_c^2}\right)^{1/2},\beta\right]}{J(\alpha,\beta)}\,
,
\eeq
where  $J(a,b)=\int_0^a (1+x^2)^{-3b/2}\,dx$ and 
$\theta_{\rm c}= R_{\rm c}/d_{\rm A}$. For the purposes of our discussion we will assume that $\beta=2/3$ and $\alpha=10$~\cite{Bartlett:00}, in which case 
\beq
J(a,2/3)=\tan^{-1}(a)\,.
\eeq
\begin{figure}[!h] 
\setlength{\unitlength}{1cm}
\centerline{\psfig{file=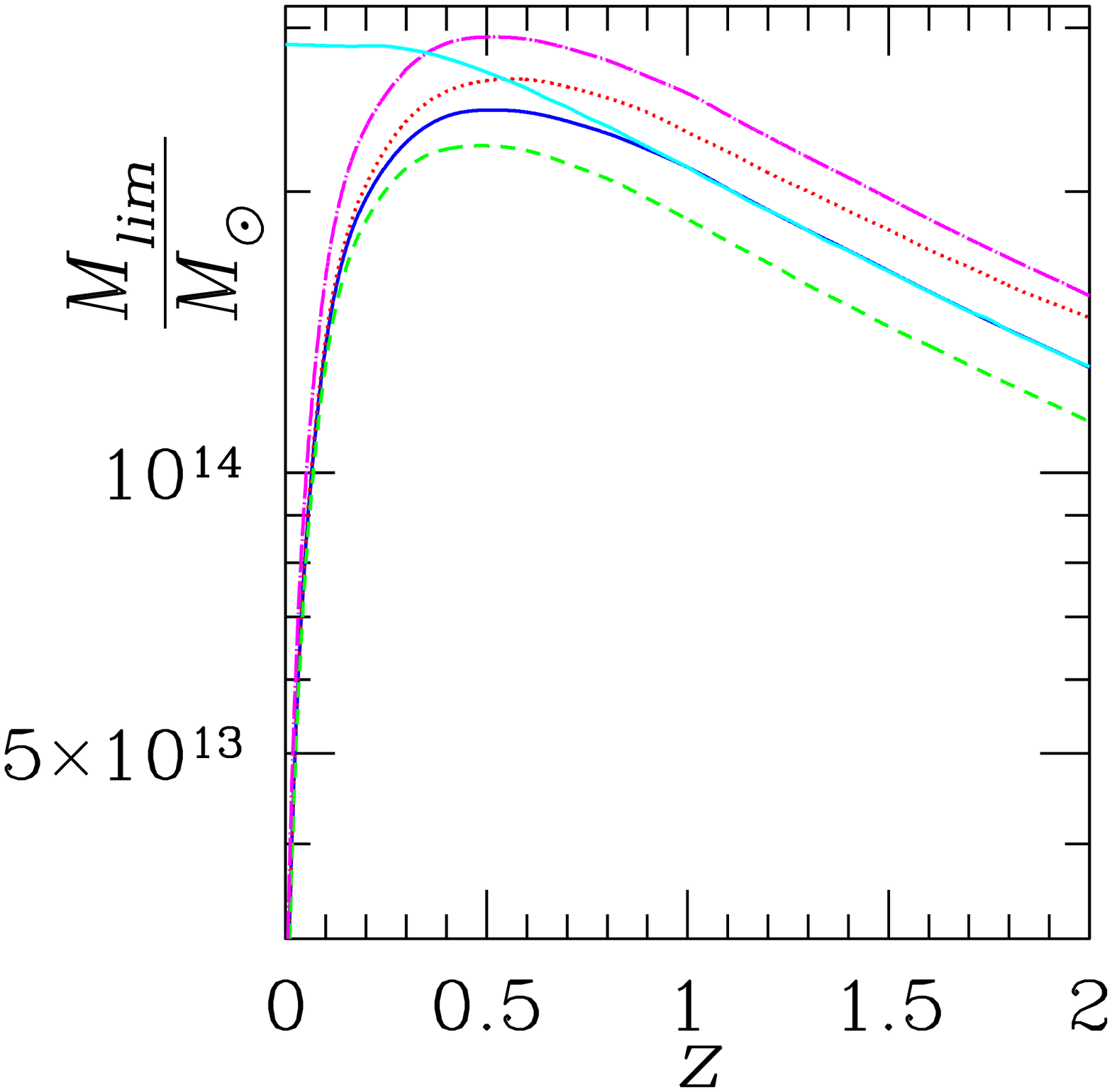,width=7cm,height=7cm}\psfig{file=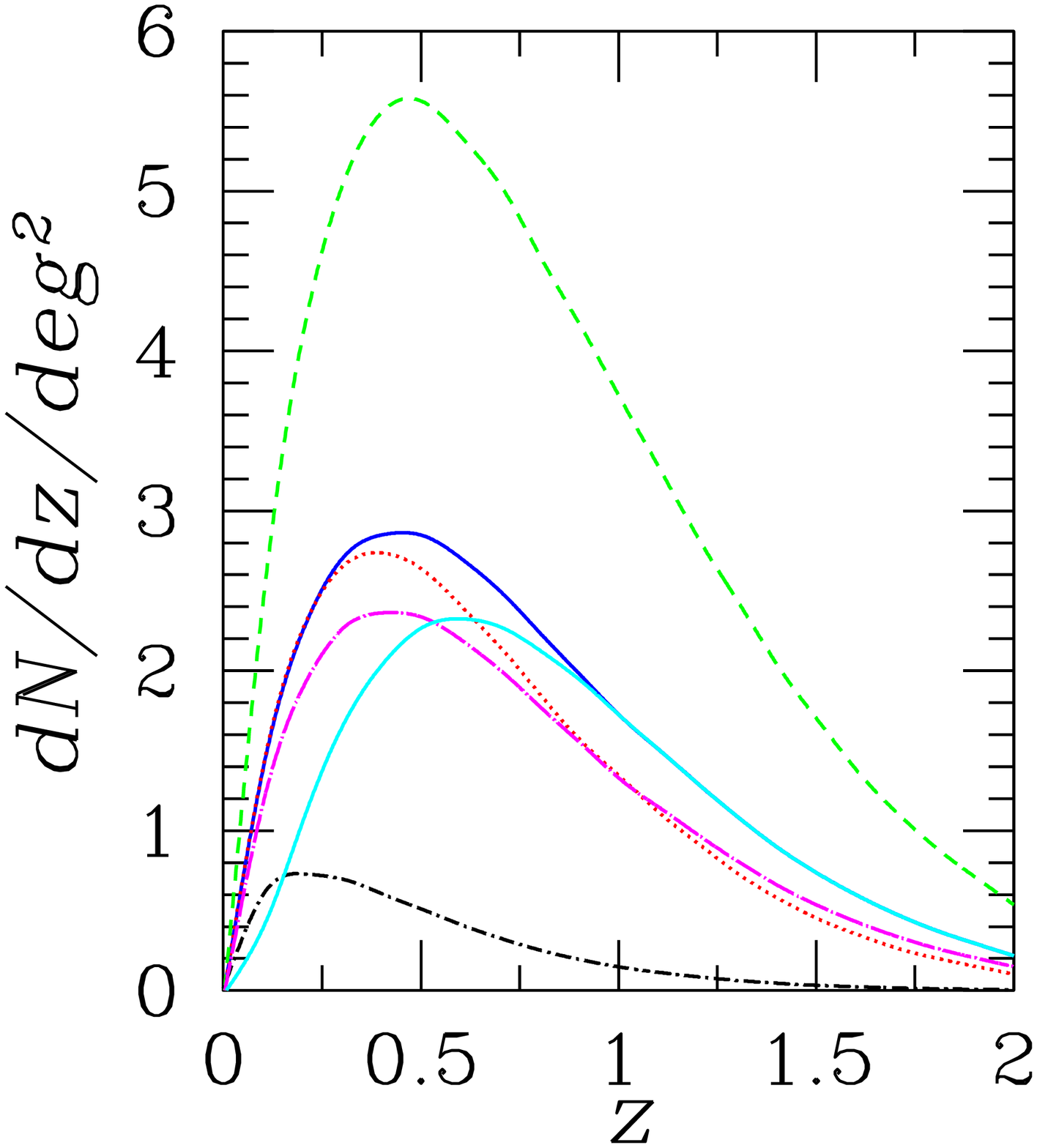,width=8cm,height=7cm}}
\caption{The limiting mass (left) and surface density of
clusters (right) for a survey with limiting flux $S_{\rm
lim}=2.62\,{\rm mJy}$ per beam, a frequency $\nu=150\, {\rm GHz}$ and a beam
width of $\theta_{\rm FWHM} = 1.3^\prime$ for an optimal combination
of pixels up to a angular size of $\theta_{\rm FWHM}^{\rm max} = 20^\prime$. The different
lines correspond to different cosmological parameters with the key the
same as in Fig.~\ref{fig:dncosmo}. The light solid line
corresponds to the {\em no} beam smoothing case, when we fix $N_{\rm b}=1$.} 
\label{fig:dnreal}
\end{figure}	
Fig.~\ref{fig:dnreal} illustrates the limiting mass and redshift
distribution of clusters if we use a beam of $\theta_{\rm FWHM}
= 1.3^\prime$. The cosmology dependence is very similar to the point
source approximation. We also include the result if  
we allow no beam smoothing (light, solid line). We see that below a
redshift of $z=1$ the possibility of beam smoothing becomes an
important feature. With $\theta_{\rm FWHM}=1.3^\prime$ 
and $\theta_{\rm FWHM}^{\rm max}=20^\prime$ it is possible to combine up to
$60$ beams, which leads for $z=0.4$  to a limiting mass which
is $5\times 10^{13} {\rm M}_\odot$ lower than without
smoothing. Furthermore we note that the limiting mass is {\rm not}
dropping to zero as it is the case in the point source approximation. For the
surface density we see how combining the pixels
improves the situation in the low redshift region. We have now all
the ingredients to perform a likelihood analysis for different types
of the surveys, which we will discuss in the next section.

\section{Likelihood Analysis of Proposed Surveys}\label{sec:likelihood}

\subsection{Proposed surveys}

In order to be able to make some firm statements as to the extent to
which one might be able to constrain cosmological parameters we have
to specify the flux limit $S_{\rm lim}$, the angular coverage
$\Delta\Omega$, the frequency of observation $\nu$ and the beam size
$\theta_{\rm FWHM}$ of the survey. This is a large parameter space and
it would be prohibitive to explore it fully. Hence, we have decided to
devote our attention to modelling six instruments which have been proposed
and which cover a wide range of possibilities. They are the
VSA~\cite{VSA}, BOLOCAM~\cite{Bolocam:1,Bolocam:2}, the Arcminute
Microkelvin Imager (AMI)~\cite{AMI:1,AMI:2}, the One-Centimetre Receiver Array (OCRA)~\cite{OCRA},
the Planck Surveyor Satellite and the South Pole Telescope
(SPT)~\cite{SPT:1,SPT:2,SPT:3}. We have not specifically include the
Sunyaev-Zel'dovich Array (SZA)~\cite{SZA:1,SZA:2,SZA:3} and the Array
for Microwave Background Anisotropy (AMIBA)~\cite{Amiba} in our
analysis since they are likely to of very similar sensitivity to
AMI. Other experiments are proposed, but we do not have sufficient information to include them in our calculations.
We should caution the reader that these instruments use a wide
range of different techniques, have/will cost wildly different amounts
of money and will become operational over a wide range of timescales. It is not our intention to discuss the merits of particular experiments rather give a picture of the likely progression of the subject.

The values which we have chosen to represent each instrument are presented in
Table~\ref{tab:surveys}. The particular values for $S_{\rm lim}$ and
$\Delta\Omega$ use a quoted instantaneous sensitivity (either
for the whole instrument or individual beams) and the optimal depth of
the survey using the procedure discussed in \cite{Battye:02}. The
angular coverage can then be computed from the instantaneous field of
view and assuming one full year integration time. Very different
results would be achieved by trading off sensitivity for area
according to $S_{\rm lim}/\sqrt{\Delta\Omega}$. However, any changes from the
quoted values would lead to fewer objects being found and would, therefore, be likely to lead to weaker constraints on cosmological parameters. This statement is only true for our particular modelling of the distribution of clusters.

It is worth making some comments on how we have done this for each of the different instruments:

\medskip

\noindent {\it VSA :} A 14-element interferometer based on
Tenerife. We use the specification for a proposed upgrade which should
have an instantaneous sensitivity of $I=0.5\,{\rm Jy}\,{\rm sec}^{1/2}$ on
a field of view of $\Omega_{\rm FOV}=1\,{\rm deg}^2$.  

\medskip

\noindent {\it BOLOCAM :} This instrument has 144 instantaneous beams
and is being used on the 8m telescope at  the Caltech Sub-millimetre
Observatory on Hawaii. We estimate $\theta_{\rm FWHM}=0.8^{\prime}$ at
150 GHz and each of the beams has $I=35\,{\rm mJy}\,{\rm
sec}^{1/2}$. Atmospheric emission could well require some beam
switching, but for our calculations we assume that all 144 beams
measure the sky providing an instantaneous field of view $
\Omega_{\rm FOV}=0.025\,{\rm deg}^{2}$. 

\medskip

\noindent {\it AMI :} Consists of the eight 13m dishes of the Ryle
Telescope and a new  array of ten 3.7m antennas. It should have
$\sqrt{\Omega_{\rm FOV}}=21^{\prime}$ and $I=20\,{\rm mJy}\,{\rm
sec}^{1/2}$. The extra resolution provided by the Ryle baselines should allow for an investigation of the potential problem of radio point sources inside the
clusters which would dilute the signal at low frequencies.

\medskip

\noindent {\it OCRA :} Currently funded is a 10 beam system which will
use 30 GHz receivers similar to those developed for the
Planck LFI system. This will be mounted on the 32m telescope at Torun
in Poland. Since the atmosphere is likely to be significant at this
location the receivers will be used in pairs each with $I=5\,{\rm
mJy}\,{\rm sec}^{1/2}$. We present results for a proposed 100 beam
system which should have $\Omega_{\rm FOV}=0.034{\rm deg^2}$. 

\medskip

\noindent {\it PLANCK Surveyor :} Here, we use the likely sensitivity
for 2 whole scans of the sky which should take 14 months for the
100 GHz channel on the HFI instrument. We conservatively assume that it
should be possible to extract the SZ signal over 1/2 of the sky. We
have not attempted to optimize the yield of this survey since the
satellite will cover the whole sky. 

\medskip

\noindent {\it SPT :} This is an ambitious project to build a 10m
telescope at the South Pole and mount on it a 1000 element bolometric
array. It should be possible for such an array to achieve $S_{\rm
lim}=5\,{\rm mJy}$ on $\Delta\Omega=4000\,{\rm deg}^2$~\cite{SPT:3} \\

\noindent The number of clusters we predict for the fiducial cosmology and the
setups of Table~\ref{tab:surveys} are
$\approx 60$ for VSA, $\approx 400$ for BOLOCAM, $\approx 170$ for
AMI, $\approx 460$ for OCRA, $\approx 4500$ for SPT and $\approx 5200$
for the Planck surveyor.

\begin{table}[!h]
\centering
\begin{tabular}{ccccc}
 & $S_{\rm lim}\;[{\rm mJy}]$ & $\nu\;[{\rm GHz}]$ & $\theta_{\rm
fwhm}$ & $\Delta\Omega\; [{\rm deg}^2]$ \\
\hline 
\hline 
VSA& $5.75$ & $30$ & $8.0^\prime$ & $300$ \\
\hline 
BOLOCAM & $1.64$ & $150$ & $0.8^\prime$ & $130$ \\
\hline 
AMI & $0.58$ & $15$ & $4.5^\prime$ & $230$ \\
\hline 
OCRA & $0.30$ & $30$ & $1.1^\prime$ & $140$ \\
\hline
Planck & $41.5$ & $100$ & $9.2^\prime$ & $20600$ \\
\hline
SPT & $2.62$ & $150$ & $1.3^\prime$ & $1430$ 
\end{tabular}
\caption{The experimental parameters for the six different SZ surveys assuming approximately one year integration time.}
\label{tab:surveys}
\end{table}

\begin{figure}[!h]
\setlength{\unitlength}{1cm}
\centerline{\psfig{file=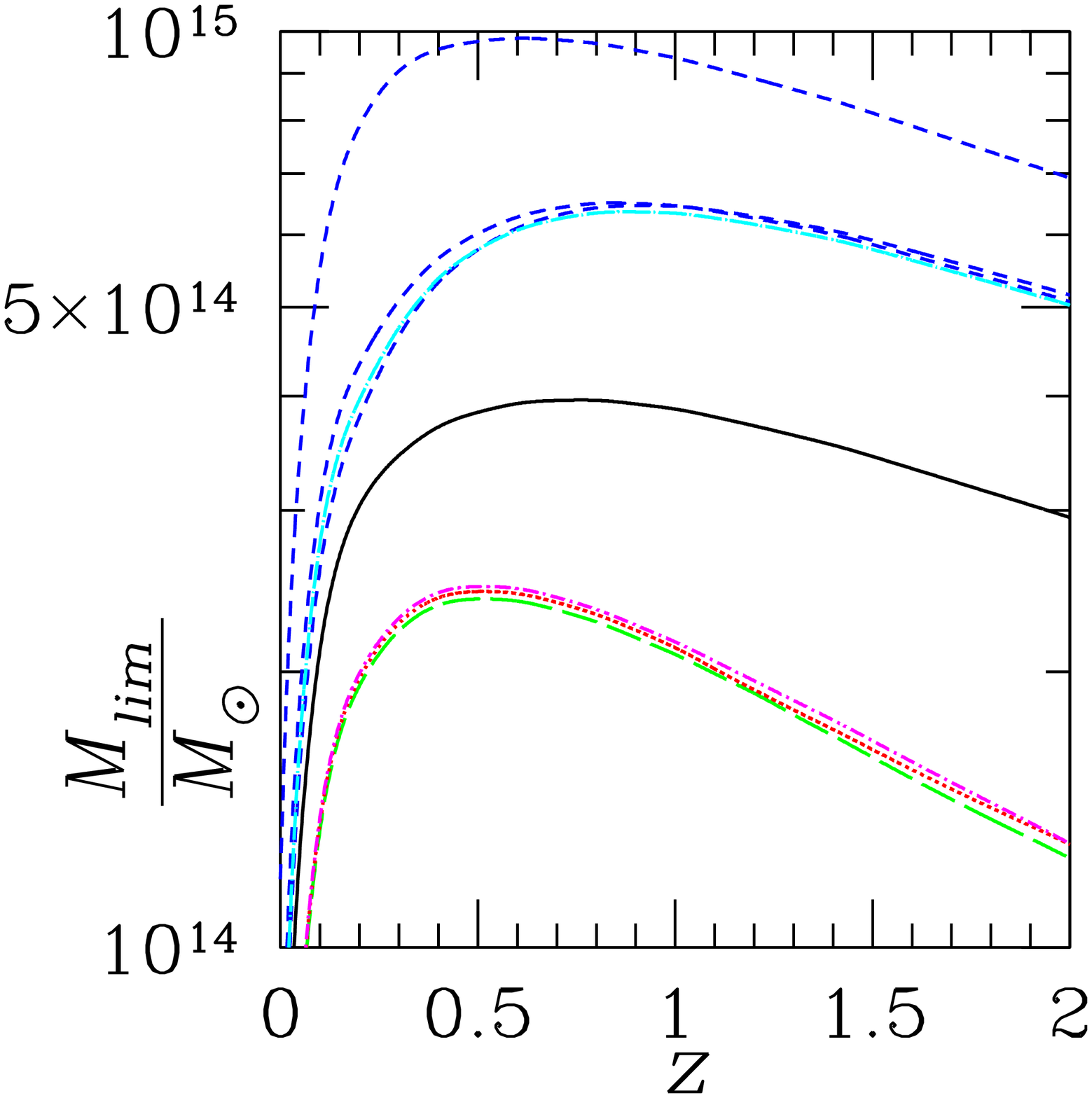,width=7cm,height=7cm}\psfig{file=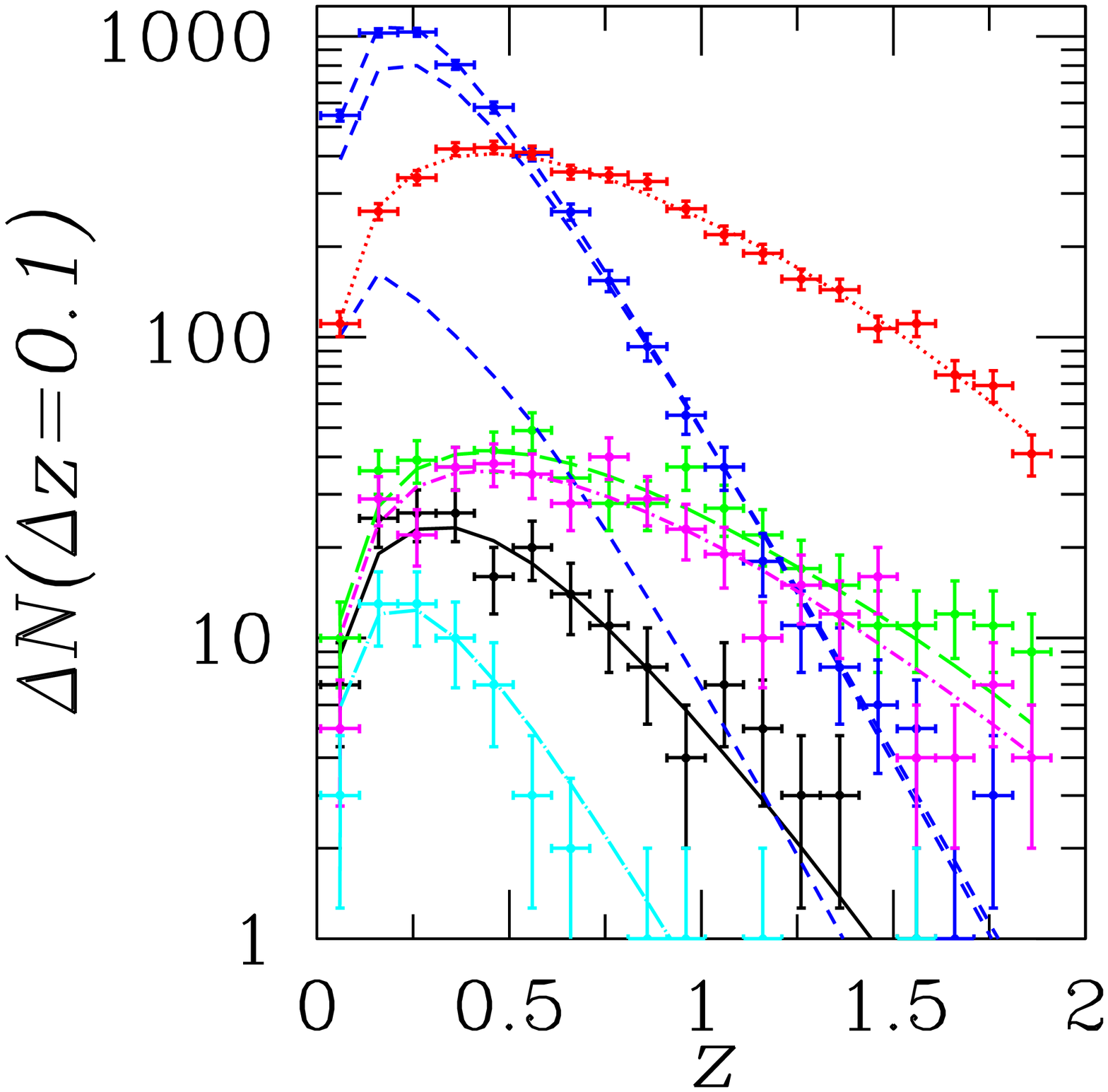,width=8cm,height=7cm}}
\caption{Limiting mass (left) and number of clusters (right) 
in redshift bins of
$\Delta z = 0.1$ for the different experimental setups from
Table~\ref{tab:surveys}. The dot - short dash line is for BOLOCAM, the
dot - long dash line for VSA, the solid line is for the AMI
survey, the long 
dashed line for OCRA, short dashed lines for the 100, 143 and 353 GHz
channels of the Planck Surveyor (note that we only use the 100 GHz
channel in the further analysis since it predicts the largest number
of clusters) and the dotted line for SPT. The data points are from a
randomly generated sample assuming 
Poisson errors. Note that we have approximated the errorbars in the plot
with a Gaussian distribution of width $\sqrt{\Delta N}$. The plots are
all for the fiducial cosmology.}
\label{fig:dnexp}
\end{figure}

In Fig.~\ref{fig:dnexp}, we present the computed mass limit of the proposed
surveys. Due to the fact that we have optimized the survey yield,
instruments with similar beam sizes have similar limiting masses, that
is  SPT, OCRA and BOLOCAM have the lowest limiting masses and are
approximately the same, as are VSA and Planck which have the highest
limiting mass because of their large beams and low sensitivity per
beam. On the right in Fig.~\ref{fig:dnexp} we show the surface density
of clusters in 
bins of $\Delta z =0.1$. VSA should find a handful of clusters in each bin
while AMI could discover up to $\approx$ 20 clusters in some bins around
the peak near $z=0.3$. BOLOCAM and OCRA will observe a few tens of
clusters in the redshift bins around the peak region and many more at
high redshifts than 
Planck. This is is due to the small beam size. Although SPT will
observe fewer clusters than Planck, in the region above redshift
$z=0.5$ the cluster yield of SPT is far larger than the one from
Planck, again due to its comparably small beam. 

\subsection{Poisson statistics}

In the following we will discuss the ability of these experiments to
constrain cosmological parameters. Therefore we need to define a
statistic to treat this problem quantitatively. We assume that the
measurement is dominated by Poisson statistics.
If we measure $N_{{\rm bin},i}$ clusters in a
particular redshift bin $z_i$ the overall probability to observe $\{n_i\}$
clusters in $N_{\rm dat}$ bins is then
\beq
    p\left(\left\{n_i\right\}|\left\{N_{{\rm bin},i}\right\}\right) =
\prod_{i=1}^{N_{\rm dat}}\frac{N_{{\rm bin},i}^{n_i}}{n_i!}e^{-N_{{\rm
bin},i}}\, .
\eeq
We would like to test how probable it is that a given theory
fits our measurement $\{N_{{\rm bin},i},z_i\}$. Therefore, we have to
calculate $p\left(\left\{N^{\rm theory}_{\rm bin}(z_i)\right\}|\left\{N_{{\rm
bin},i}\right\}\right)$, with $N^{\rm theory}_{\rm
bin}(z_i)=\int_{z_i-\Delta z/2}^{z_i+\Delta z/2} dN/dz(z)\, dz$. We
can build a log-likelihood statistic by defining 
\beq
	C = -2 \ln p = -2 \sum_{i=1}^{N_{\rm dat}} N^{\rm theory}_{\rm
bin}(z_i)
\ln N_{{\rm bin},i} - N_{{\rm bin},i} -\ln N^{\rm theory}_{\rm bin}(z_i)\, ,
\label{eqn:llike}
\eeq
which is known in the literature as Cash C statistic~\cite{Cash:79}. We
will sample the log-likelihood function in (\ref{eqn:llike}) around
our fiducial cosmology, where we assume that we can get redshifts out
to $z_{\rm max}=1.5$ with an 
accuracy of $\Delta z=0.1$. This
should be feasible with photometric redshifts from instruments like
the SDSS and VISTA~\cite{SDSS:1,SDSS:2,VISTA}. For
an efficient way of sampling we adopt a Markov chain Monte Carlo
sampling method~\cite{Christensen:00,Christensen:01,Lewis:02}, where
we typically compute half a million of samples to achieve
convergence. 
For the likelihood scanning we vary $H_0$, $\Om$, $\sigma_8$, $w_0$ and $w_1$
and keep the other parameters fixed. In
order to speed up the likelihood calculation we also approximate the
calculation of $\sigma(R)$ in (\ref{eqn:sig}) by the expressions
given in \cite{Sugiyama:95,Viana:99}.

\begin{figure}[!h]
\setlength{\unitlength}{1cm}
\centerline{\psfig{file=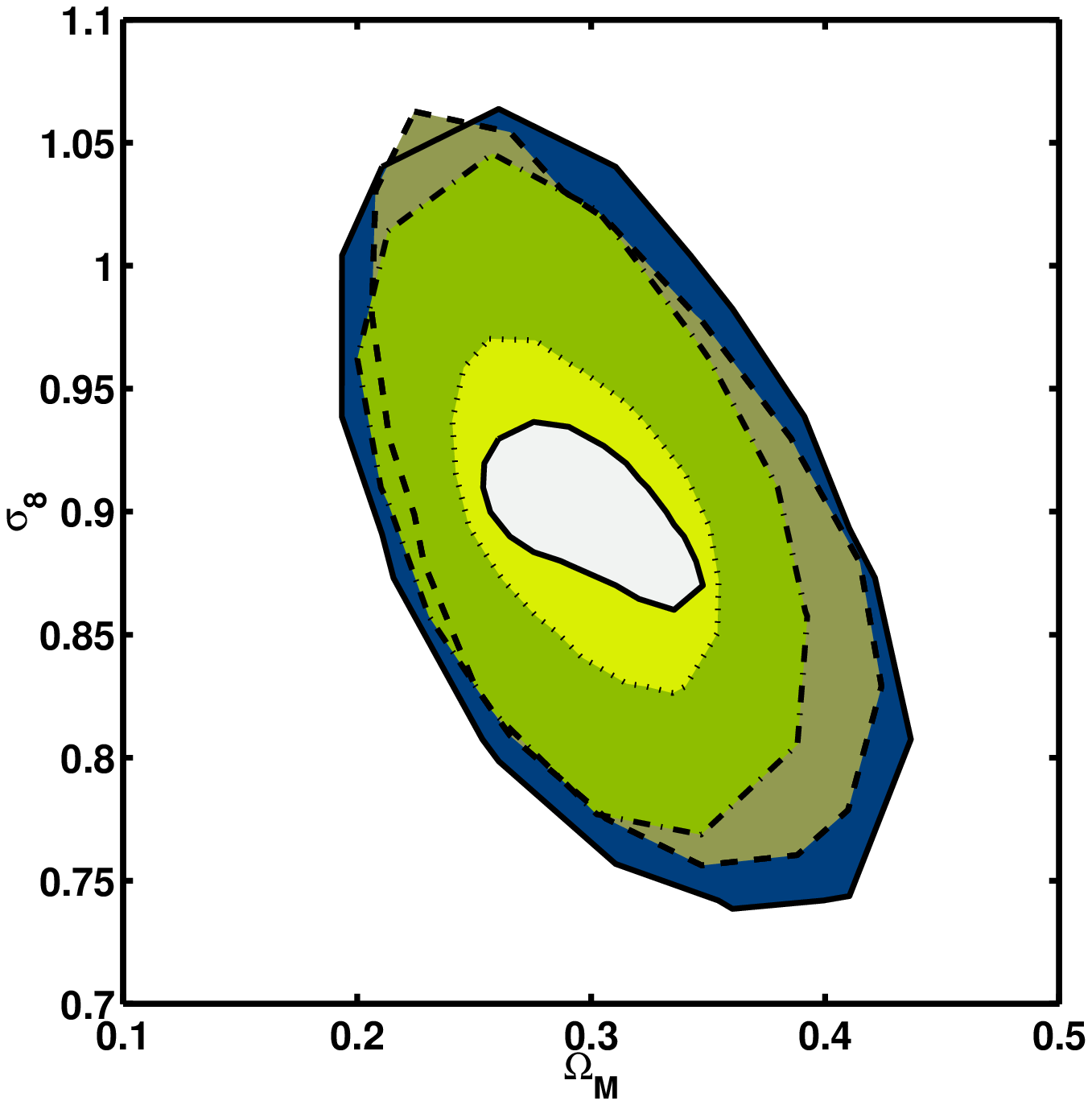,width=6cm,height=6cm}\psfig{file=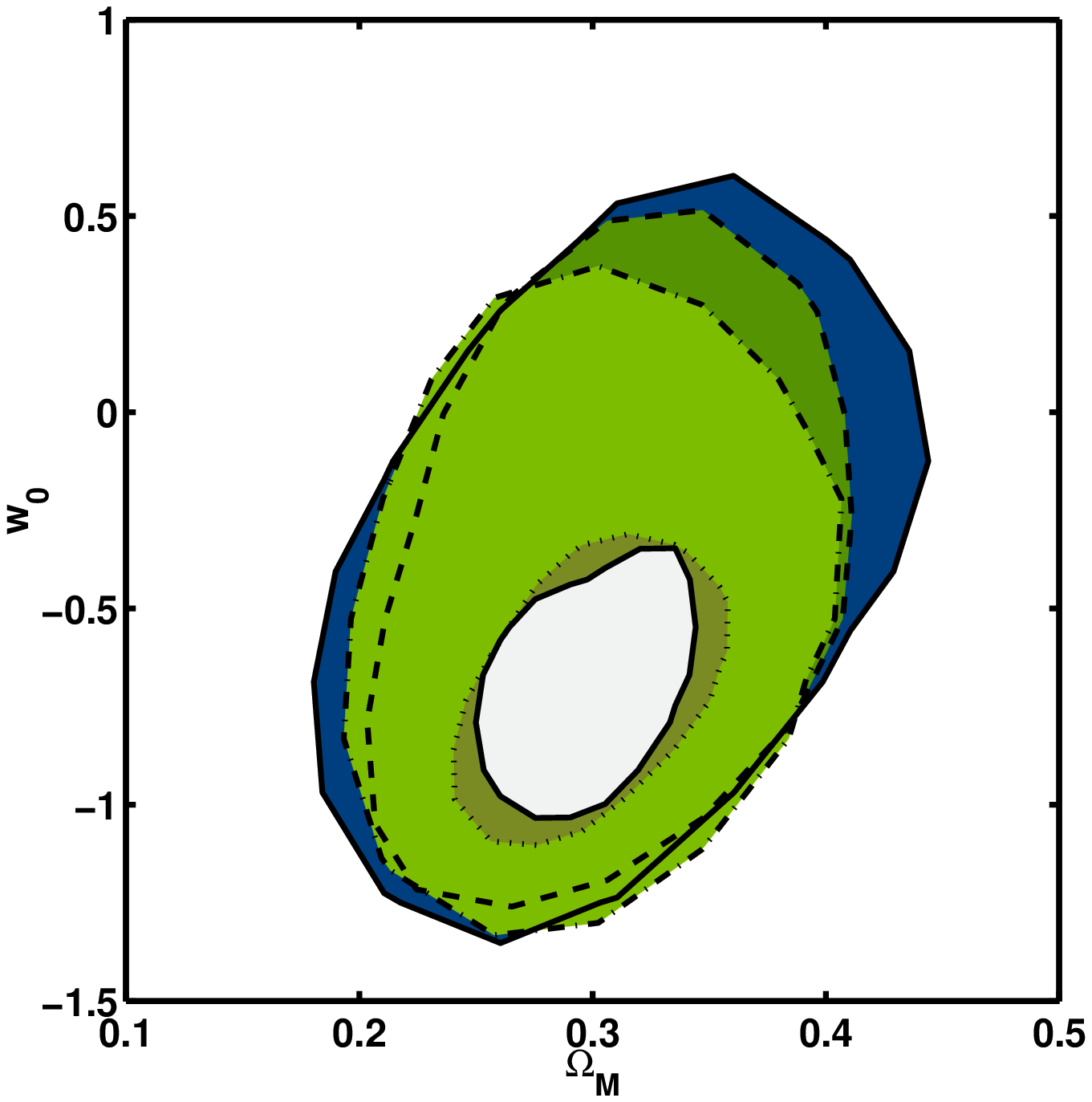,width=6cm,height=6cm}\psfig{file=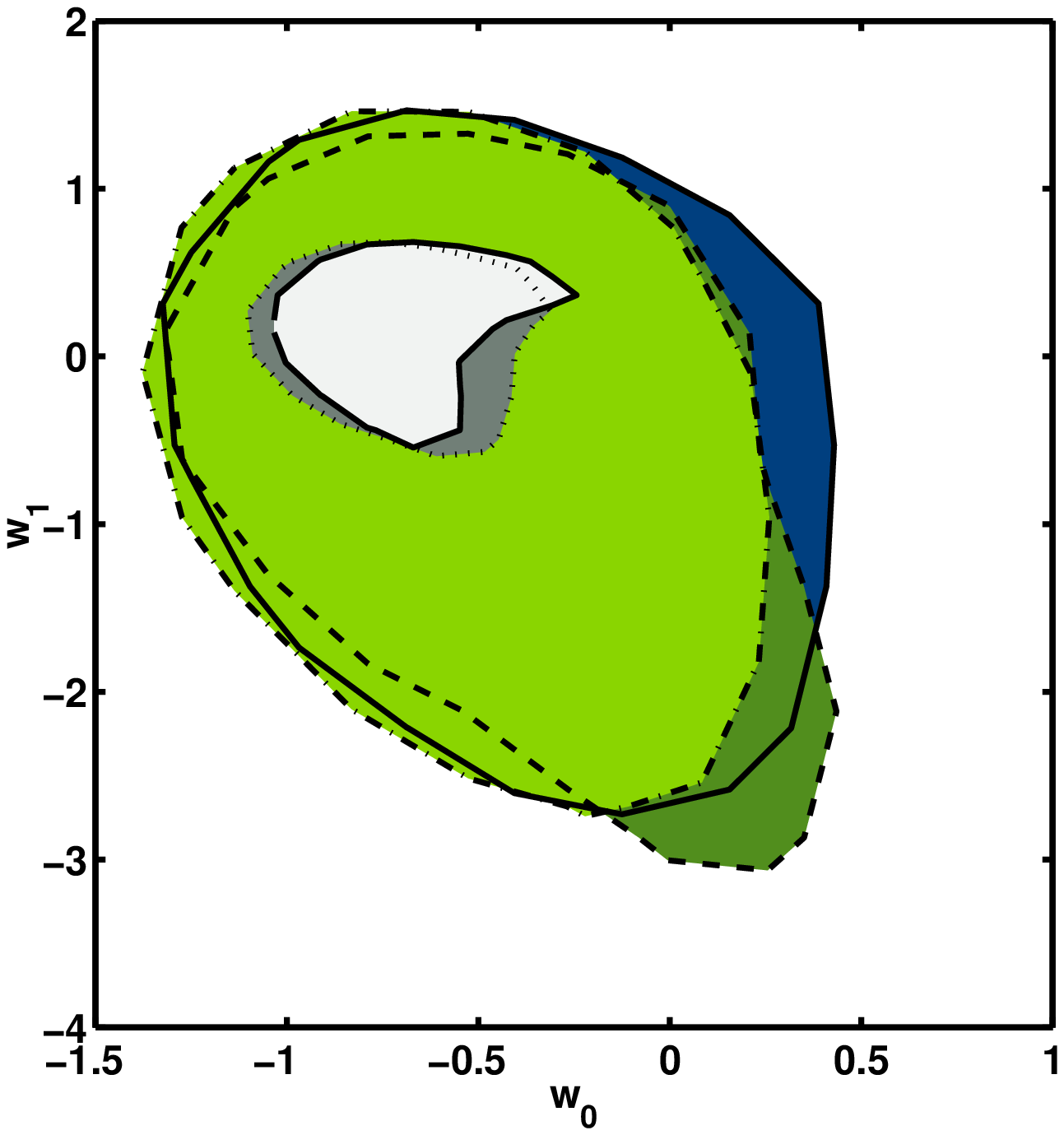,width=6cm,height=6cm}}
\caption{$1-\sigma$ joint likelihood contours marginalized over all
the other parameters with a prior on $H_0=72\pm 8\, {\rm
km}\,{\rm sec}^{-1}\,{\rm Mpc}^{-1}$.
From outside in the contours correspond to AMI, BOLOCAM,
OCRA, SPT and Planck. Note that VSA yields weaker 
constraints than AMI and has been omitted.}
\label{fig:likelihood}
\end{figure}

\subsection{Mock likelihood analysis}

In Fig.~\ref{fig:likelihood} we present the joint $1-\sigma$ likelihood
contours for the mock surveys. We have marginalized over all the other
parameters and the only prior assumption made was that $H_0 = 72\pm
8\, {\rm km}{\rm sec}^{-1}{\rm Mpc}^{-1}$ as suggested by the Hubble
Space Telescope Key Project~\cite{Freedman:01}. Note that we include a
larger number of parameters in our analysis than in previous analyses
which naturally tends to increase the size of the errorbars. It is
clear that the results from VSA, BOLOCAM, AMI and OCRA are unlikely to
improve on current constraints on the cosmological parameters, but
that they will provide extra independent, complementary information
which can be used in conjunction with, for example, that from the
primary CMB. It is possible that these surveys could be used in
conjunction with X-ray or weak lensing observations of clusters to
probe the gas physics of clusters and as well as to constrain
cosmological parameters, providing further insight for future more
powerful surveys. 

Our analysis shows that Planck and SPT are likely to yield
tight constraints on $\sigma_8$ and $\Omega_m$. They will be also able
to start to constrain the equation of state parameters. However,it
seems that the surface density of clusters alone will not be able to
distinguish the fiducial model from a $\Lambda$CDM cosmology. In order
to achieve tighter constraints on the equation of state it will
necessary to use additional complementary information from SNe
observations~\cite{SNAP}, baryon mass fraction
measurements~\cite{Allen:02a} or other cluster abundance
measurements~\cite{Allen:02b}. In the subsequent discussion we will
focus on the SPT survey. 

\begin{figure}[!h]
\setlength{\unitlength}{1cm}
\centerline{\psfig{file=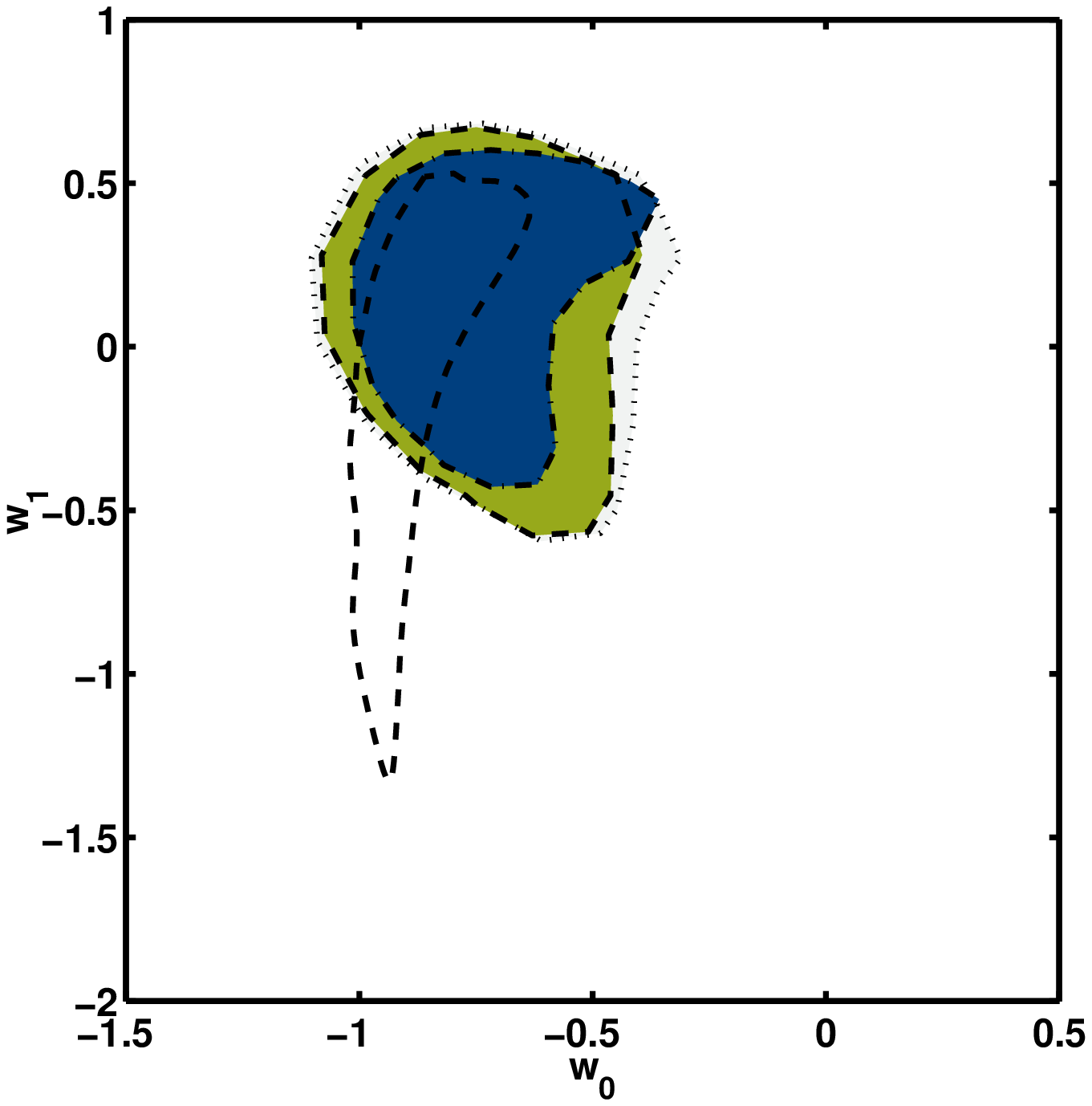,width=6cm,height=6cm}\psfig{file=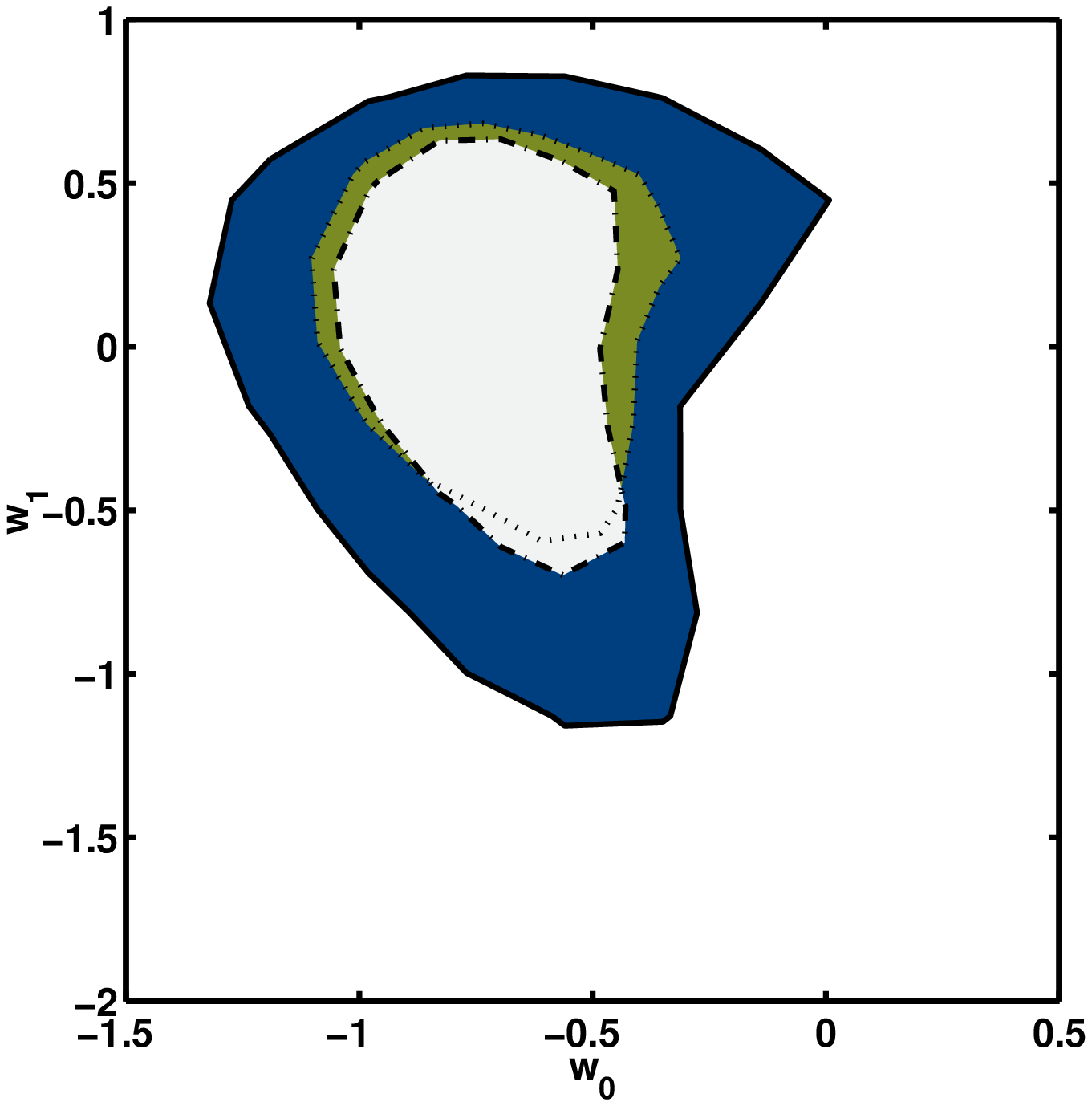,width=6cm,height=6cm}} 
\caption{The $1-\sigma$ joint likelihood in the $w_1-w_0$ plane for
the SPT survey. On the left, standard SPT (dotted), with
a prior on $\Omega_m=0.3\pm0.04$ (dashed filled contour) and with
prior on $\Om=0.3 \pm 0.02$ and $\sigma_8=0.9 \pm0.01$ (inner
dot-dashed filled contour). The dashed line is for the Supernovae
Acceleration Probe -- SNAP.  On the right, $z_{\rm max}=1.0$ (solid), standard SPT (dotted),
SPT with $\Delta z=0.01$ (dot-dashed).}    
\label{fig:likelihood_par}
\end{figure}

In Fig.~\ref{fig:likelihood_par}, we investigate the use of
prior information on  $\Om$ and $\sigma_8$. First, we include an
uncertainty of $\Delta \Om 
= \pm 0.04$ (solid). This accuracy has been already 
achieved for measurements of the baryonic gas fraction in clusters using X-ray techniques~\cite{Allen:02a}. It was established that this
uncertainty is relatively insensitive to the inclusion of a dark energy
component and hence we use it without modification~\cite{Bridle:03}. We
see that such a prior does not significantly improve our ability to
constrain the dark energy parameters. In order to improve on this, we
have also put very tight priors on $\Om$ {\em and} $\sigma_8$, with
$\Delta \Om = 0.02$ and $\Delta \sigma_8 = \pm 0.01$ (dot-dashed); a level of
constraint which could be possible using X-ray and lensing
observations~\cite{Allen:02b}. This serves illustrate how sensitive 
the cluster abundance measurement is to the value of $\sigma_8$ since the constraint on the dark energy is substantially improved. If
there is a tight prior on $\sigma_8$ then other cosmological parameters can be tightly constrained using SZ surveys.

The dashed open contour is the
$1-\sigma$ likelihood for the proposed SNAP survey. We assumed for
this analysis that SNAP will discover about 2000 SNe out to redshift
of $z=1.7$~\cite{Weller:01,Weller:01a}. We see that  SZ surveys
provide valuable complementary information to SNe surveys. The SNe
seem to constrain the constant part of the equation of state, $w_0$,
much more tightly than the evolving part $w_1$, while SZ clusters
appear to constrain both parameters equally. This is because the
surface density of clusters essentially constrains cosmological
parameters via the linear growth factor $D(z)$ and the volume element
$dV/(dzd\Omega)$, while SNe constrains on the magnitude-redshift
relation. As a note of caution we find that  even using SPT with the
tightest priors as well as SNAP one  cannot establish the evolution
of the equation of state of the fiducial model. However, if we
marginalize over $w_1$, a  difference between cosmological constant
and $w_0>-1$ can be established albeit with limited significance. 

In Fig.~\ref{fig:likelihood_par} (right) we illustrate how our
knowledge of the cluster redshifts can affect our conclusions. The
outermost solid line likelihood contour is for the SPT survey with
$z_{\rm max}=1$. In this case, the errorbars are much larger than
those expected when we assume the  measurement of redshifts out to
$z_{\rm max}=1.5$. The dashed contour is obtained if we have redshift
bins of $\Delta z=0.01$ instead of  
$\Delta z = 0.1$. We see that there is very little difference
between the error contours for these two different binning values
suggesting that photometric redshifts should be sufficient at least
for this application. This can be seen by looking back at
Fig.~\ref{fig:dnexp} where we see that $\Delta z = 0.1$ seems
sufficient to map the shape of the surface density of 
clusters since it evolves much more slowly with redshift. 

\section{Statistical and Systematic Uncertainties}\label{sec:systematics}
\subsection{Mass - temperature relation}\label{sec:mtsys}
In order to calculate the surface density of clusters above a
certain mass limit it is necessary 
assume a  mass-temperature relation (\ref{eqn:mt}). The
derivation of the mass temperature relation assumed  that clusters are
completely virialized objects in thermal equilibrium governed only by
gravitational physics. However on-going mergers, incomplete
virialization and numerous possible non-gravitational heating
mechanisms, such as AGNS and SNe,  can modify this assumption. We
will, therefore, discuss the  consequences of modifications to the
mass-temperature relation. First we will concentrate on the overall
normalization amplitude 
$T_*$ which we have already pointed out is deduced by normalization to
the results of  numerical simulations. From the compilation of
different simulations and  
observations in~\cite{Huterer:02a} we see that $T_*$ can be anywhere in the
range $T_*=1.0-2.2$. We have chosen $T_*=1.6$ for our analysis which is in
the middle of this region, although not preferred by a particular
measurement or simulation.

\begin{figure}[!h] 
\setlength{\unitlength}{1cm}
\centerline{\psfig{file=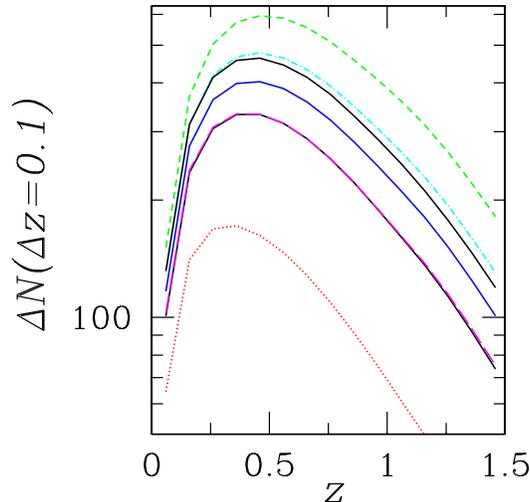,width=7cm,height=7cm}}
\caption{The surface density 
for the SPT survey for different values
of $T_*$. The dotted line is for $T_*=1.0$, the long dashed line (partially hidden) for
$T_*=1.44$ ($-10\%$), the dot-dashed line for $T_*=1.76$
($+10\%$) and the dashed
line for $T_*=2.0$. The middle solid line is for the fiducial
model, the upper solid line for $\Om=0.32$ and the lower solid line
for $H_0=65\, {\rm km}\,{\rm sec}^{-1}\,{\rm Mpc}^{-1}$. Clearly,
differences of $\pm 10\%$ are of the same order of magnitude as the
constraints on cosmological parameters which we are trying to establish.} 
\label{fig:dN_bh1}
\end{figure}

In Fig.~\ref{fig:dN_bh1}, we show the variation of the redshift
distribution of clusters with varying $T_*$. We also show the
values for the fiducial model and models with
$\Om=0.32$ and $H_0=65\, {\rm km}\,{\rm sec}\,{\rm Mpc}^{-1}$.
We see from this that an uncertainty in $T_*$
changes the amplitude in the surface density but {\em not} the
shape and, therefore, we expect that the degeneracy between the equation of
state parameters, which mainly defines the shape of the curves (see
Fig.~\ref{fig:dnreal}), is largely unaffected by this
uncertainty. However, we expect degeneracies with $\sigma_8$ and $\Om$.

\begin{figure}[!h]
\setlength{\unitlength}{1cm}
\centerline{\psfig{file=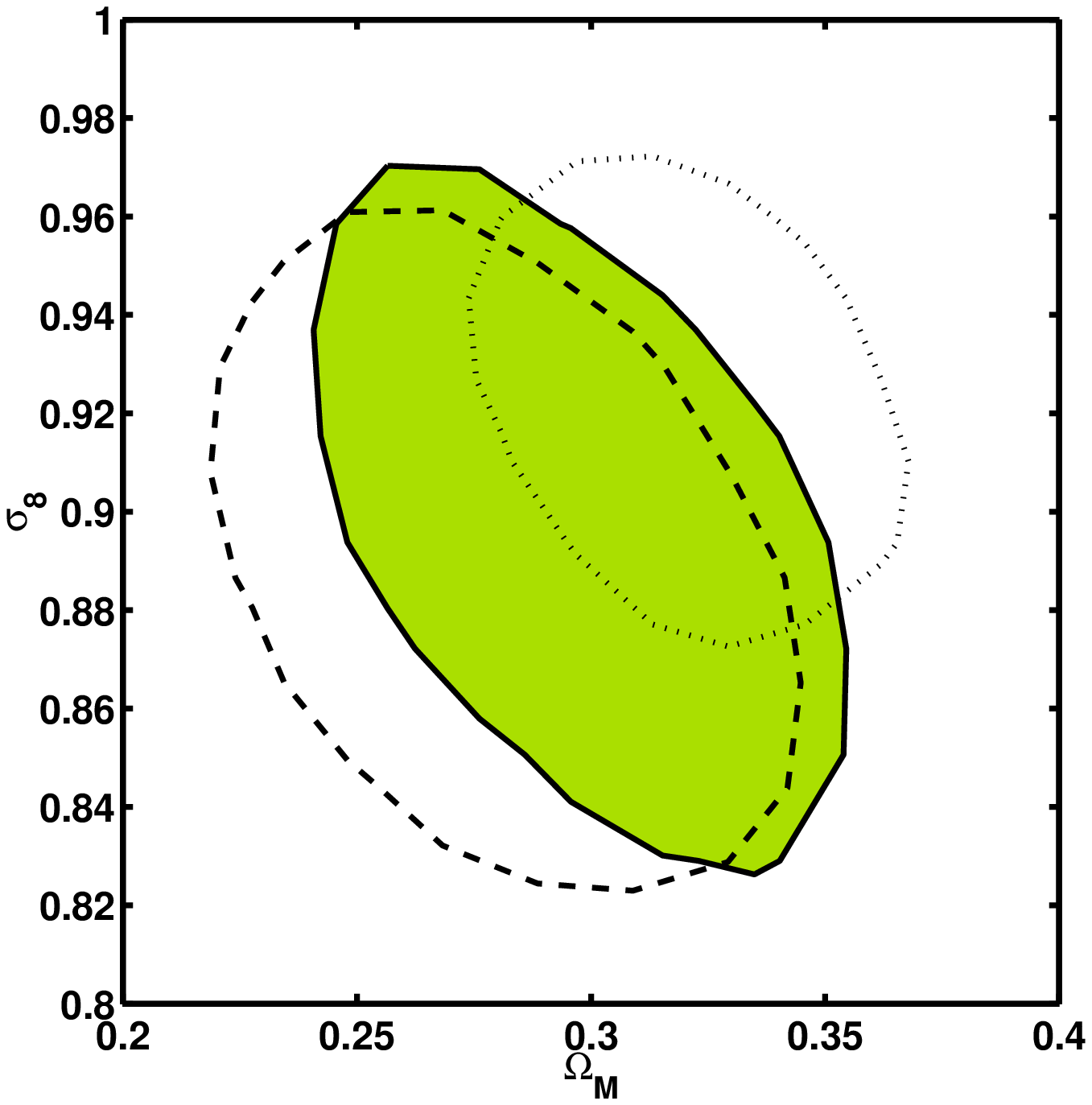,width=6cm,height=6cm}\psfig{file=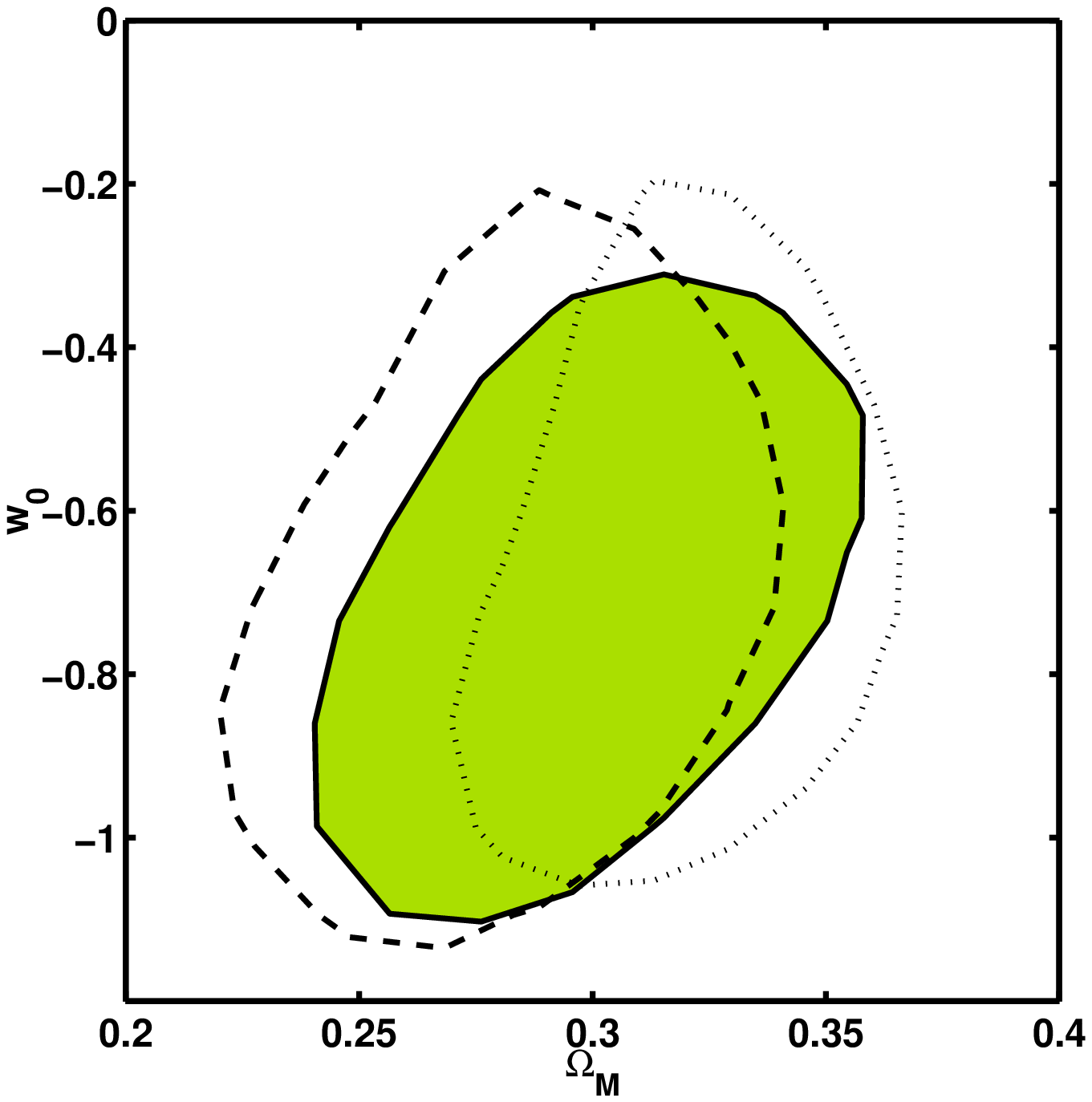,width=6cm,height=6cm}}
\centerline{\psfig{file=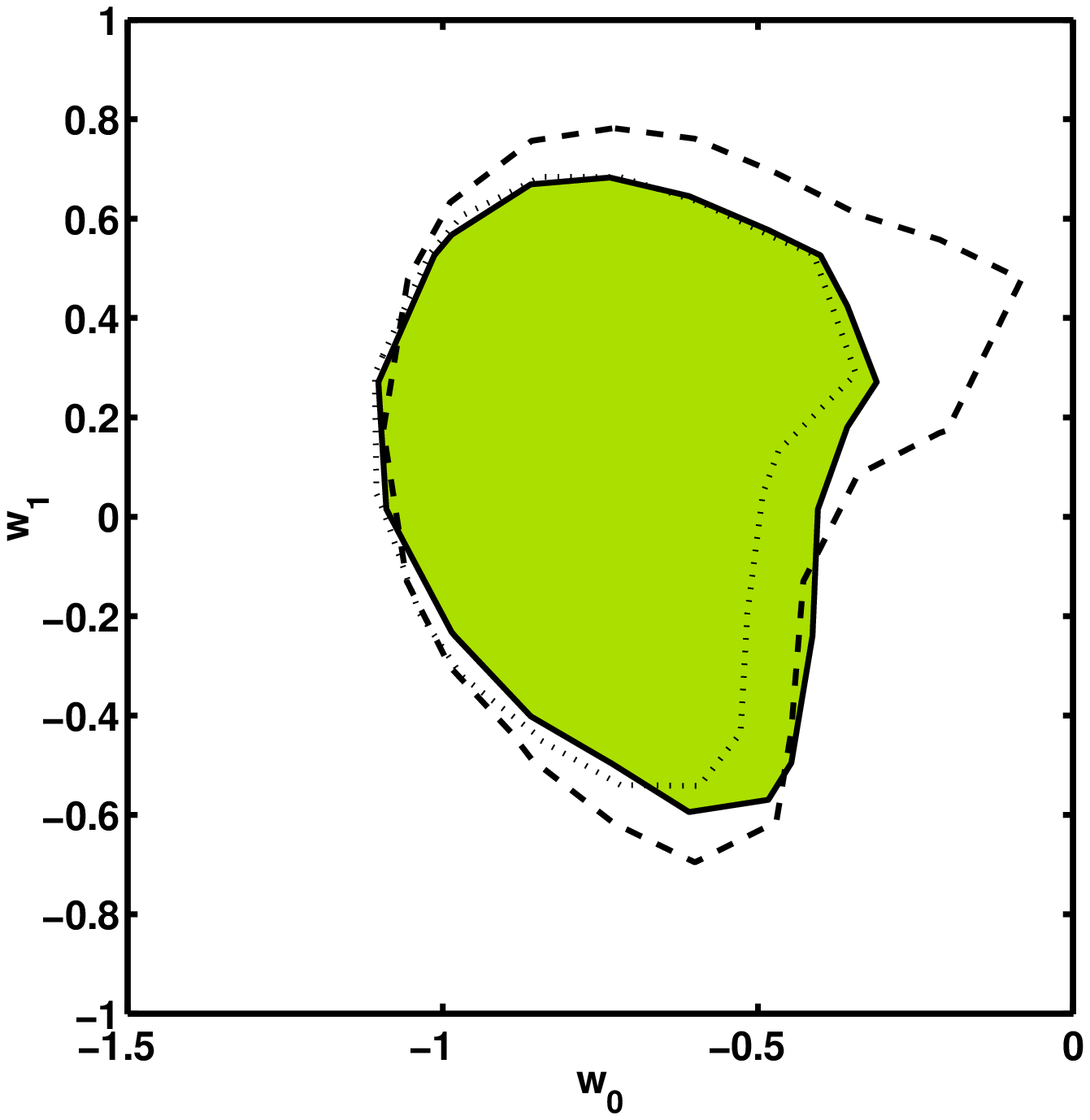,width=6cm,height=6cm}\psfig{file=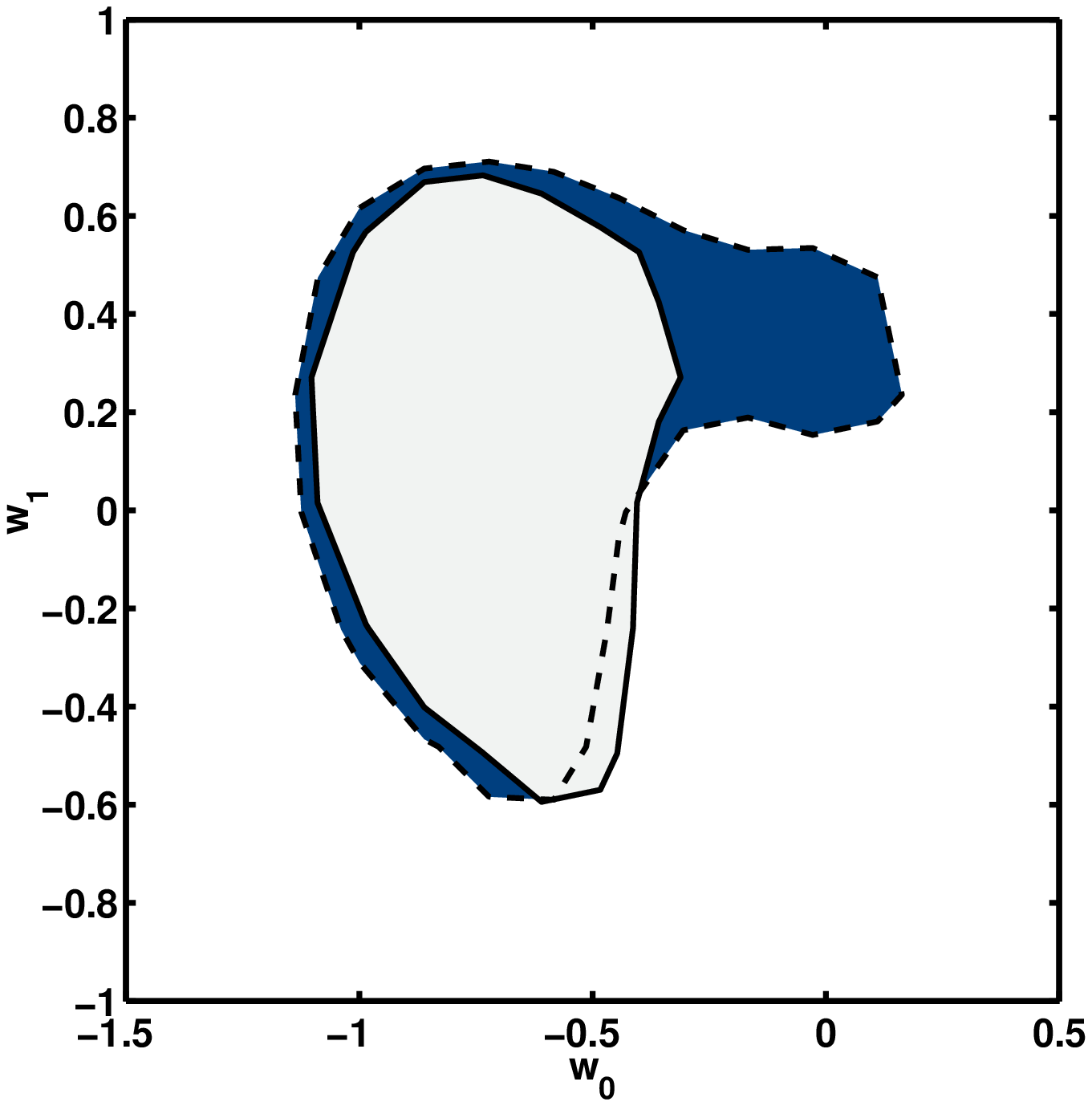,width=6cm,height=6cm}}
\caption{The $1-\sigma$ likelihood contours for the SPT survey. In the
top two panels and the lower left panel we show the effect of the
mass-temperature normalization bias, if we assume $T_*$ is $16\%$
larger (dotted) and $12\%$ lower (dashed) than the fiducial value of
$T_*=1.6$. In the lower right panel $T_*$ is included as a
free parameter. The inner light shaded contour is with fixed $T_*=1.6$
and the outer contour marginalized over $T_*$.}
\label{fig:dN_bh2}
\end{figure}

In order to test this, we first study the bias we introduce
if we analyse mock data, created using a different value of $T_*$ than
the one we used to perform the analysis. In the top two and lower left
panel of Fig.~\ref{fig:dN_bh2} we show the results of this. The
dotted line corresponds to a model created with $T_*=1.86$ and the
dashed line to $T_*=1.42$. In all cases we performed the analysis for
the fiducial value $T_*=1.6$. This roughly corresponds to a bias in
the limiting mass of $\pm 20\%$. We see clearly the  bias introduced in the $\sigma_8$ and $\Om$ measurements by choosing the wrong value of $T_*$. However, the effect on the dark energy parameters $w_0$ and
$w_1$ (Fig.~\ref{fig:dN_bh2}, lower left) seems to introduce only a slight extra
bias toward larger $w_0$ values and only increases or lowers the
errorbars corresponding to the change in the overall number of
observed clusters. 

This becomes even clearer if we include the parameter $T_*$ in the 
likelihood analysis and marginalize over $T_*$. The result of
including $T_*$ is presented in Fig.~\ref{fig:dN_bh2}, lower right
panel. We see that this  only increases the errorbars in the $w_0$
direction. This is 
because $w_1$ (and also partly $w_0$) influence mainly the shape of
the surface density and not so much its amplitude, as suggested above.

So far we have assumed that the mass-temperature relation applies
universally. However, at this stage it is only an observed correlation
and is likely to have considerable scatter. Next, we investigate the
inclusion of statistical uncertainties in the limiting mass. We
incorporate this via a ``selection function''
$\phi(M,z)$~\cite{SPT:1}, 
\beq
	\frac{dN}{dz} =
{\Delta\Omega}\frac{dV}{dzd\Omega}(z)\int\limits_0^\infty \phi(M,z)\frac{dn}{dM}\,dM\, ,
\eeq
with 
\beq
	\phi(M,z) = \frac{1}{2}\left\{{\rm
erf}\left[\frac{M-M_{\rm lim}(z)}{\delta M_{\rm lim}(z)}\right]+1\right\}\, ,
\eeq
where $\delta$ is the relative statistical uncertainty in the mass
limit $M_{\rm lim}$. 
\begin{figure}[!h] 
\setlength{\unitlength}{1cm}
\centerline{\psfig{file=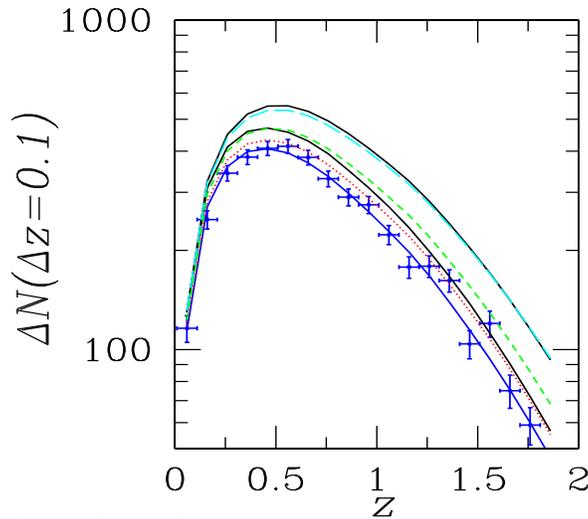,width=8cm,height=7cm}}
\caption{The surface density of clusters 
for the SPT survey. The
lowest solid line is for the fiducial model, the middle solid line for
$\Om=0.32$ and the upper solid line for $\sigma_8=0.95$. The dotted
line is for a statistical uncertainty in the limiting mass of
$\delta=20\%$, the short dashed line for $\delta=30\%$ and the long
dashed line for $\delta=40\%$. The data points are from a simulated
mock catalogue for the SPT survey.}
\label{fig:dN_delta}
\end{figure}
In Fig.~\ref{fig:dN_delta} we illustrate the influence of different
values of $\delta$ on the observed surface density. 
A general result is that the inclusion of this effect using a symmetric
selection function will always increase the overall number of observed
clusters since there are many more objects just below the mass
threshold than just above it.  
We see from the figure that an unknown
uncertainty in the limiting mass will also increase the uncertainty
in constraints on the cosmological parameters, but if it is known and
well understood it should be possible to incorporate it into the
analysis.  
Since the changes in the surface density due to $\delta\ne 0$ only
become larger than the expected Poisson errors for SPT when the
scatter is greater than $\delta=15\%$, we estimate that a statistical 
uncertainty below this level should be acceptable for this survey. The
Poisson errors for the other surveys are much larger and hence it
should be possible for them to withstand much larger values of
$\delta$ and still give sensible constraints. 

The final uncertainties in the mass temperature relation which we
shall investigate are a different power law dependence between $M$ and
$T$,  and a different overall dependence on
redshift~\cite{Verde:01,SPT:1}. We re-define the mass temperature
relation to be 
\beq
\left<T_{\rm e}\right>_n = T_*\left(\Delta_c
E(z)^2\right)^{1/3}\left[1+\left(1+3w\right)\frac{\Oq(z)}{\Delta_c}\right]\left(1+z\right)^{\varepsilon-1}\left(\frac{\Mv}{10^{15}h^{-1}M_\odot}\right)^{1/\xi}\,
,
\eeq
where if $\varepsilon=1$ and $\xi=3/2$ we obtain the standard mass
temperature relation (\ref{eqn:mt}). Note that new values for
$\varepsilon$ and $\xi$ would require a recalibration of the
relation and hence a different value of $T_*$. However, since we will
only restrict ourselves to a  qualitative discussion we will not incorporate
this effect. The parameter $\xi$ could model non-gravitational heat input.
Since smaller clusters (that is, groups) are preferentially affected by these processes, we expect $\xi\ge1.5$. Observations and
simulations suggest values of $1.48\le\xi\le1.98$
~\cite{Mohr:97,Mohr:99,Muanwong:01,Xu:01,Finoguenov:01}. The parameter
$\varepsilon$ models deviations from complete virialization. On one hand,
clusters at early times might not be
completely virialized, hence $\varepsilon<1$. However, clusters which
have ongoing mergers or some other form of energy injection could be
much hotter than expected and hence have $\varepsilon>1$. At this stage there is no observational preference for any particular value of $\varepsilon$.
\begin{figure}[!h]
\setlength{\unitlength}{1cm}
\centerline{\psfig{file=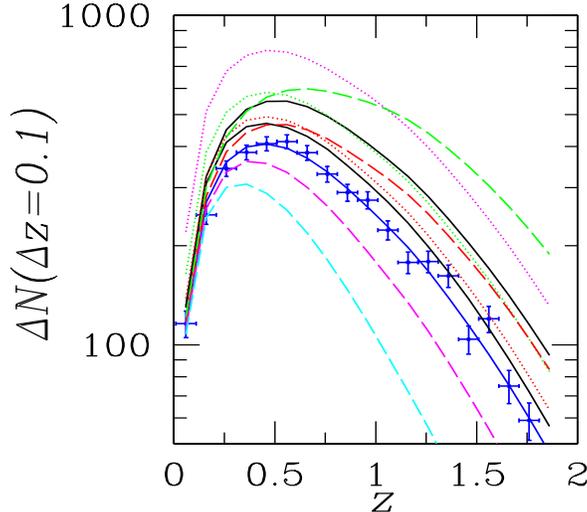,width=8cm,height=7cm}}
\caption{The surface density of clusters 
for the SPT survey with
changes in the mass - temperature relation. The
lowest solid line is for the fiducial model, the middle solid line for
$\Om=0.32$ and the upper solid line for $\sigma_8=0.95$. The dotted
lines are for a change in the power law between limiting mass and
temperature. The lowest dotted line is for 
$\xi=1.6$, the middle dotted line for $\xi=1.7$ and the top for $\xi=1.9$.
The long dashed lines are for changes in the redshift dependence of
the $M-T$ relation. Form top to bottom the long dashed lines
correspond to $\varepsilon=1.5,1.2,0.8,0.5$. The data points are from a simulated
mock catalogue for the SPT survey.}
\label{fig:dN_xi}
\end{figure}
In Fig.~\ref{fig:dN_xi} we illustrate the surface density 
for the SPT
survey for different values of $\xi$ and $\varepsilon$. We allow $\xi$ to vary
in the range $\xi=1.6,1.7,1.9$ and see that we observe more
clusters for higher values of $\xi$. This is clear since from
(\ref{eqn:finalflux}) we obtain $\Mv \propto
S_\nu^{\xi/(\xi+1)}$. This again will increase the uncertainties
mainly in the parameters which define the amplitude of the redshift
distribution, like $\Om$ or $\sigma_8$. We vary
$\varepsilon$ between $0.5$ and $1.5$. A
recent Fisher matrix analysis of SZ cluster counts for the SPT survey
obtained uncertainties of $\Delta\xi \approx
0.0064$ and $\Delta\varepsilon \approx 0.46$~\cite{SPT:1}, if they are
included as parameters. Although we 
believe that a Fisher matrix analysis can only give crude errorbars,
we expect degeneracies to not restrict our ability to extract  
the cosmological parameters. We should note that follow up
measurements which constrain the mass of the cluster independently
using, for example, X-ray measurements or weak lensing, could
constrain the non-standard mass 
temperature relation considerably~\cite{SPT:1,Verde:01} and, hence,
improve the veracity of constraints on the cosmological parameters deduce from SZ surveys.

\subsection{Mass - temperature relation and $\sigma_8$}

Recent years have seen a wide range of values reported in the
literature for the power spectrum normalization
$\sigma_8$. Observations with 
weak lensing, X-rays, Large-Scale-Structure and the large scale  SZ
effect suggest values in the range
$0.61\le\sigma_8\le1.05$~\cite{Pierpaoli:02}. If interpreting a
particular observations requires the conversion 
from temperature to mass, as is the case for the SZ effect and X-ray
observations, then it is important to know the value of $T_*$ which
has been used, since there is a direct degeneracy 
between $\sigma_8$ and the normalization $T_*$ with $\Om^{0.6}\sigma_8
\propto T_*^{-0.8}$~\cite{Pierpaoli:02}. As noted in the introduction
one problem of our analysis is that the choice of a lower value for  
$\sigma_8$ in our fiducial model would have resulted in a lower number of
observable clusters and hence much weaker constraints on the
cosmological parameters due to the Poisson statistics. However, the
analyses of X-ray measurements which result in $0.7\le\sigma_8\le0.8$
use a larger value of
$T_*$~\cite{White:02,Huterer:02,Pierpaoli:02}. We will now discuss
various aspects of this issue. 
\begin{figure}[!h] 
\setlength{\unitlength}{1cm}
\centerline{\psfig{file=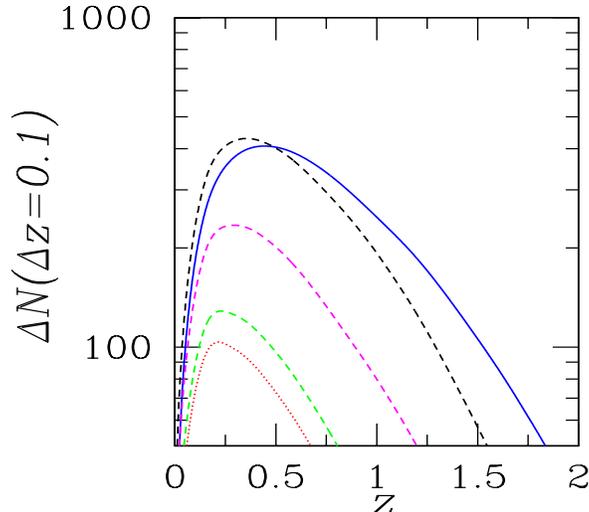,width=8cm,height=7cm}}
\caption{The surface density of clusters 
for the SPT survey. The
solid line is for the fiducial parameters and set up. The dotted line
is for the same setup, but $\sigma_8=0.72$. The dashed lines are for
the $\sigma_8=0.72$ with $T_*$ increasing from $1.8$ (lowest
dashed line) via $T_*=2.5$ (middle dashed line) to $T_*=3.5$ (top
dashed line).}
\label{fig:dNdz_mt}
\end{figure}
In Fig.~\ref{fig:dNdz_mt} we illustrate  the consequences of combining lower
values of $\sigma_8$ with larger values of $T_*$. The solid line is
for the fiducial model and the SPT survey. The dotted line is for the
same set up with $\sigma_8=0.72$, which seems to be in the preferred range
of X-ray and Large-Scale-Structure
observations~\cite{Lewis:02,Allen:02a,Allen:02b}. We see that the overall
number of clusters is considerably less for the lower value of
$\sigma_8$. However, X-ray observations seem to require a lower
calibration of the mass temperature relation 
~\cite{Allen:02b,Pierpaoli:02}, which corresponds to  an increased value of 
$T_*$. In Fig.~\ref{fig:dNdz_mt} we show the increasing number of clusters observed for
as $T_*$ increases. Still for the marginally maximal acceptable value of
$T_*=2.5$~\cite{Huterer:02a} the number of clusters is about a factor
of $1.7$ below the number observed in a universe with
$\sigma_8=0.9$. Nonetheless it is clear that the lower values of
$\sigma_8$ suggested by some lead to the very small number of clusters one
would naively assume. 

In order to obtain as many clusters predicted in the fiducial model,
but using  $\sigma_8=0.72$ it is necessary to increase to $T_*=3.5$,
which corresponds 
to a $75\%$ lower normalization of the mass - temperature
relation, outside the range suggested by simulations and measured in
reality~\cite{Huterer:02a}. Hence, if one is drawn to conclusions of
these particular observations it seems likely that our fiducial model
over-predicts the number of clusters by at least a factor of two and the
errorbars on the parameters we have predict are optimistic. However,
there are observations which suggest the opposite. Preliminary
observations of secondary CMB anisotropies of the large-scale SZ
effect using the CBI instrument~\cite{CBI:SZ} indicate a value of
$\sigma_8 > 1$~\cite{Komatsu:02}. Although this results remains to be
confirmed, it indicates that the issue of the value of $\sigma_8$ is
far from settled.  
We should note that it might be possible to measure
the normalization of the mass - temperature relation directly with the
combination of cluster abundances and weak lensing
observations~\cite{Huterer:02a} and constrain both $\sigma_8$ and
$T_*$ directly. Furthermore, a 
measurement of the flux of each cluster will enable one to perform an
internal calibration of the sample and alleviate the uncertainty in the
normalization of the mass - temperature relation~\cite{Hu:03}.

\subsection{Mass function}
A further uncertainty we need to consider is the mass function
itself. In our analysis we used the 2002 results from the VIRGO
consortium~\cite{Evrard:02}. In order to obtain insight into the
uncertainty of the mass function we have compared our results to the 
 previously released mass function from 2001~\cite{Jenkins:01}
\beq
\frac{dn}{dM}\left(z,M\right) = - 0.316
\frac{\rho_m(t_0)}{M}\frac{d\sigma_M}{dM}\frac{1}{\sigma_M}
\exp\left\{-|0.67-\log\left[D(z)\sigma_M\right]|^{3.82}\right\}\, ,
\label{eqn:conum2}
\eeq
with the mass defined by an overdensity of $\Delta = 324$ relative to
the matter density, which corresponds to an overdensity of $\Delta=97$ relative
to the critical density for the fiducial universe.
We have also compared our results with the standard PS
mass function~\cite{Press:74}
\beq
\frac{dn}{dM}\left(z,M\right) =
-\sqrt{\frac{2}{\pi}}\frac{\rho_m(t_0)}{M}\frac{\delta_c}{D(z)\sigma_M^2}\frac{d\sigma_M}{dM}\exp\left(-\frac{\delta_c^2}{2D(z)^2\sigma_M^2}\right)\, 
\label{eqn:PS}
\eeq
with $\delta_c=1.686$, where we ignore the weak cosmology dependence
of $\delta_c$ in the present discussion. 
\begin{figure}[!h]
\setlength{\unitlength}{1cm}
\centerline{\psfig{file=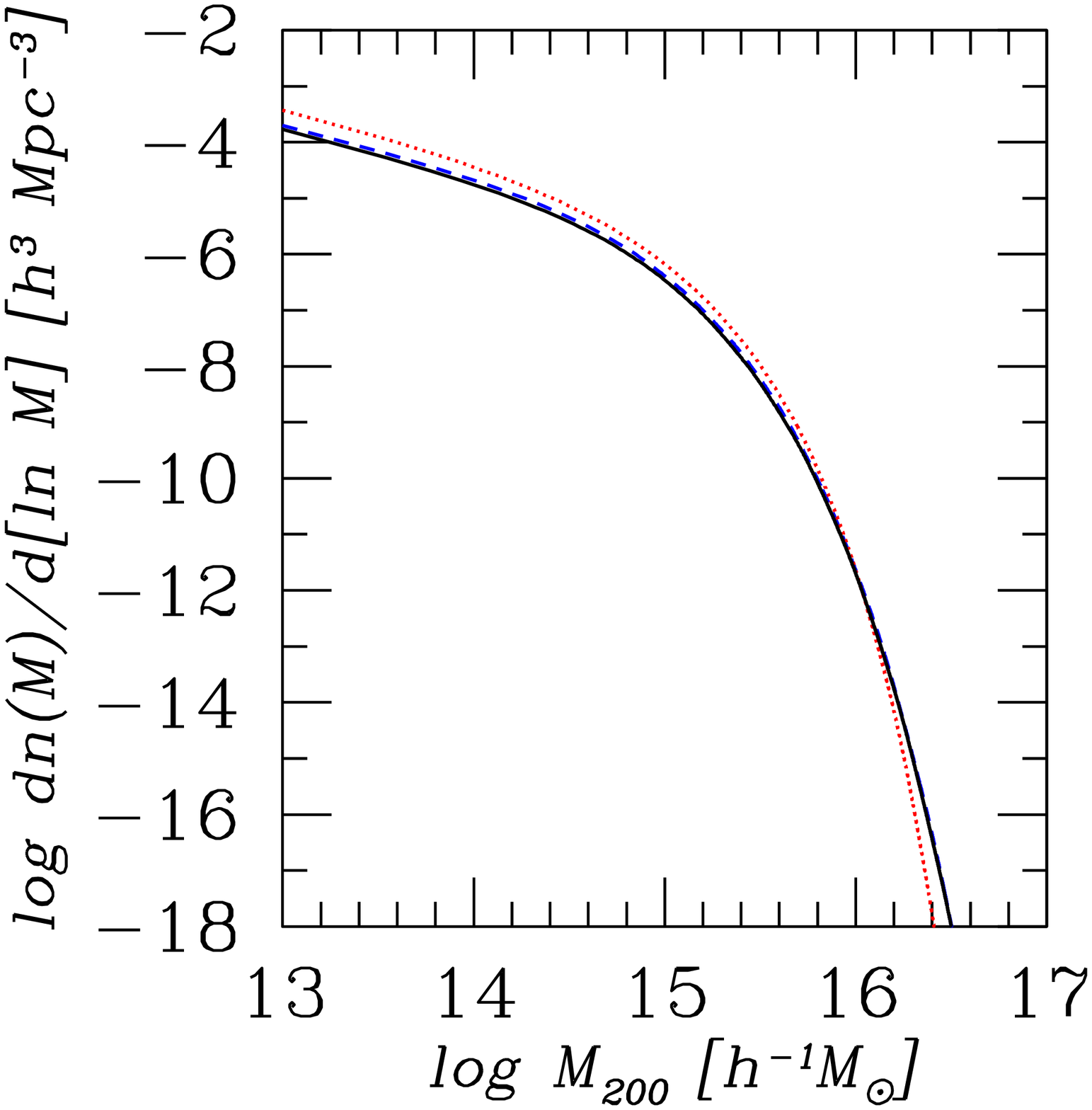,width=8cm,height=7cm}\psfig{file=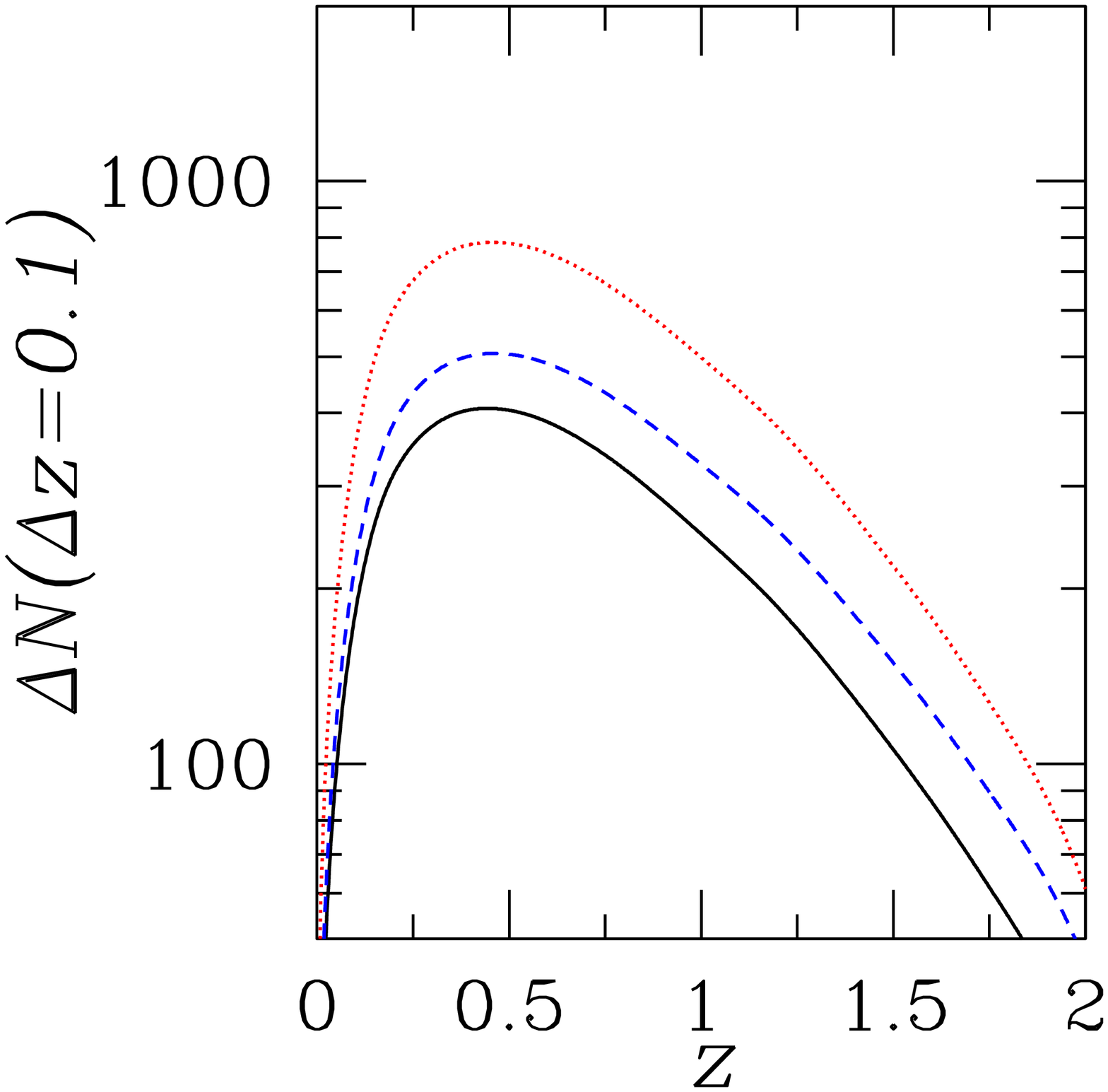,width=8cm,height=7cm}}
\caption{On the left the different mass functions at $z=0$. The dashed
line is for the Jenkins 2001 mass function~\protect\cite{Jenkins:01}, the solid line for the Evrard 2002 mass function~\protect\cite{Evrard:02} both from  the VIRGO consortium. The dotted line is
that expected from the PS formalism for a mass definition of
$M_{200}$. On the right the expected surface density 
of clusters for these mass functions for the
SPT survey.}
\label{fig:dNdz_mass}
\end{figure}
In Fig.~\ref{fig:dNdz_mass}, we show the different mass
function at $z=0$ for the fiducial cosmology. The solid line is for
the 2002 mass function of the VIRGO consortium~\cite{Evrard:02}, the
dashed line for 2001~\cite{Jenkins:01} and the dotted line for the
PS mass function. We will investigate the consequences of these three
mass functions for expected yield of the SPT survey. Since SPT has a
limit of roughly $1.7\times10^{14} h^{-1}M_\odot$ we have to consider the mass
function above this limit.The PS function dominates over both the
VIRGO 2001 and VIRGO 2002 mass 
function up to masses of $10^{16}h^{-1} M_\odot$ and then PS mass function
drops below both VIRGO mass functions. The VIRGO 2001 function is
marginally above the VIRGO 2002 function in the range below
$10^{16}h^{-1} M_\odot$. However, above this region the
number of clusters is so sparse that the contribution to the
surface density of clusters is not significant. We also present the 
surface density for the SPT survey in Fig.~\ref{fig:dNdz_mass}. We see
that the PS mass function results in about twice as many clusters 
and the VIRGO 2001  about $25\%$ more clusters than the VIRGO 2002
mass function. It is clear from this analysis, that a precise
convergence of the mass function is required to constrain cosmological
parameters using the 
surface density of clusters, providing further impetus for an extended program of numerical simulations. Clearly the uncertainty in the mass
function is degenerate with other uncertainties in the mass - temperature
relation and $\sigma_8$. However, combining  SZ, X-ray and weak lensing of clusters could help constrain the mass function as well.

\section{Discussion and conclusions}
In this paper we have introduced a realistic model for the
surface density of clusters that would be observed by an SZ instrument 
including effects of the beam of the survey and the profile of the cluster. In the Section~\ref{sec:likelihood} we analysed constraints on cosmological parameters that might be expected from such surveys. We found that SPT and Planck will be particularly powerful in constraining  the
standard parameters $\sigma_8$ and $\Omega_m$, and can be used in conjunction  with other observations, for example SNe surveys, to constrain the dark energy. These surveys can be expected on a timescale of around 5 to 7 years. In the meantime less powerful surveys such as 
those possible using VSA, BOLOCAM, AMI and OCRA will provide useful
complementary constraints, while allowing the study of the clusters
themselves, in particular using X-ray follow up. 

We have shown in Fig.~\ref{fig:likelihood_par} that it is
sufficient to use redshift bins of $\Delta z=0.1$ to constrain the
cosmological parameters. Essentially the shape of the redshift
distribution is mapped out adequately with this bin width. It should be 
feasible for surveys like SDSS or VISTA~\cite{SDSS:1,SDSS:2,VISTA} to measure redshifts of this accuracy, although one will have to be careful to avoid introducing extra selection effects.

While Planck, SPT and OCRA will constrain the  standard cosmological
parameters significantly, it will be difficult for them to provide
information about the equation of state of the dark energy
component. While they do not constrain it by themselves, they are
complementary to SNe observations as illustrated in 
Fig.~\ref{fig:likelihood_par}. SZ surveys essentially constrain
$\Omega_m$ and provide a prior for SNe surveys like the SNAP
mission. However, they become considerably more powerful if one includes a
tight prior on the amplitude of the density fluctuations $\sigma_8$,
which could be provided from complementary large scale structure
observations.

\begin{figure}[!h]
\setlength{\unitlength}{1cm}
\centerline{\psfig{file=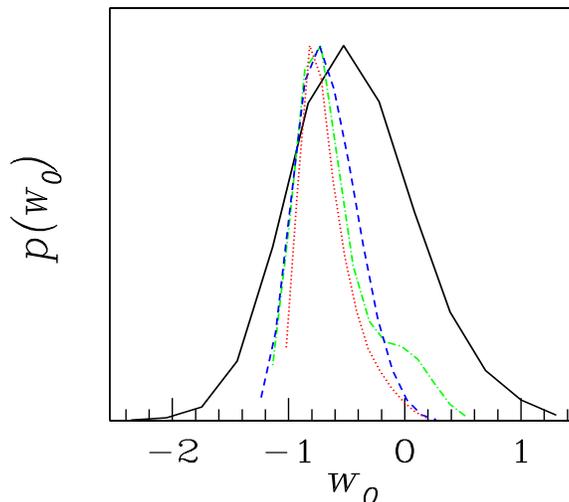,width=8cm,height=7cm}}
\caption{The probability for $w_0$ marginalized over all the other
parameters. The solid line is for the OCRA survey, the short dot-dashed
line for SPT, the dashed line for SPT including the parameter
$T_*$ in the likelihood analysis and the dotted line for SPT with
tight priors on $\Om$ and $\sigma_8$.}
\label{fig:probw0}
\end{figure}

This leaves us with the question of whether any of the proposed SZ surveys might be able to distinguish the fiducial cosmology from a $\Lambda$CDM universe.
In Fig.~\ref{fig:probw0} we show the one dimensional likelihoods for
the constant part of the equation of state $w_0$. We see that the OCRA survey can not
constrain the equation of state. However the marginalized likelihood
of SPT narrows in on the fiducial value of $w_0=-0.8$,
while with tight priors on $\Om$ and $\sigma_8$ we can distinguish the fiducial model
with $w=-0.8+0.3z$ from a $\Lambda$CDM cosmology.  

We have also discussed in detail systematic and statistical uncertainties which might affect more ability to affect our ability to extract cosmological parameters. The main sources of uncertainty are  the mass-temperature
relation and the mass function; clearly much more work is needed to
pin these two quantities down. The inclusion
of the normalization factor $T_*$ in the likelihood analysis only marginally 
increases the errorbars on the equation of state state as shown
in Fig.~\ref{fig:probw0}. A potentially more significant problem is a modified  evolution or power law of the mass temperature relation. The Fisher matrix
analysis presented in \cite{SPT:1} suggests that this problem is not too
severe. This is because the cosmological dependence of the surface density is mainly due to the growth factor, rather than that from the limiting 
mass.  

The surveys discussed in this paper will provide much more
information than just the surface density. In principle it should be
possible to use the flux information as well as the surface density to
constrain cosmological parameters as hinted in
\cite{Verde:01,Hu:03}. However, using the flux as well as redshift
information is likely to be more susceptible to systematic error. This
subject is currently under investigation. 

Finally, we should note that it is difficult to strongly constrain the
equation of state of the dark energy component, and particularly its
redshift evolution, using any single method. It is clear that a
joint effort will be required with a combination of measurements
such as those from the  cosmic microwave background, large scale
structure, SNe, weak lensing and clusters being necessary to achieve this goal.
The results presented in this paper show that the SZ effect is likely to play a strong role in this over the next decade.

\section*{Acknowledgements}
We would like to thank S.~Allen, S.~Bridle, I.~Browne, C.~Dickinson,
G.~Efstathiou,  
R.~Kneissl, J.~Mohr, J.~Ostriker and P.~Wilkinson for helpful discussions. We
also thank A.~Lewis for useful discussions and providing us with the
software to analyse the probability distributions. The parallel computations were done at the UK National Cosmology Supercomputer Center funded by
PPARC, HEFCE and Silicon Graphics / Cray Research. JW is supported by
the Leverhulme Trust and King's College, Cambridge and RAB is funded by PPARC.

\def\et{{\em et al.}}
\def\jnl#1#2#3#4#5#6{\hang{#1, {\it #4\/} {\bf #5}, #6 (#2).}}
\def\jnltwo#1#2#3#4#5#6#7#8{\hang{#1, {\it #4\/} {\bf #5}, #6; {\it
ibid} {\bf #7} #8 (#2).}} 
\def\prep#1#2#3#4{\hang{#1, #4.}} 
\def\proc#1#2#3#4#5#6{{#1, in {\it #3 (#4)\/}, edited by #5,\ (#6).}}
\def\book#1#2#3#4{\hang{#1, {\it #3\/} (#4, #2).}}
\def\jnlerr#1#2#3#4#5#6#7#8{\hang{#1 [#2], {\it #4\/} {\bf #5}, #6. {Erratum:} {\it #4\/} {\bf #7}, #8.}}
\def\prl{Phys.\ Rev.\ Lett.}
\def\pr{Phys.\ Rev.}
\def\pl{Phys.\ Lett.}
\def\np{Nucl.\ Phys.}
\def\prp{Phys.\ Rep.}
\def\rmp{Rev.\ Mod.\ Phys.}
\def\cmp{Comm.\ Math.\ Phys.}
\def\mpl{Mod.\ Phys.\ Lett.}
\def\apj{Ap.\ J.}
\def\apjl{Ap.\ J.\ Lett.}
\def\aap{Astron.\ Ap.}
\def\cqg{Class.\ Quant.\ Grav.} 
\def\grg{Gen.\ Rel.\ Grav.}
\def\mn{MNRAS}
\def\ptp{Prog.\ Theor.\ Phys.}
\def\jetp{Sov.\ Phys.\ JETP}
\def\jetpl{JETP Lett.}
\def\jmp{J.\ Math.\ Phys.}
\def\zpc{Z.\ Phys.\ C}
\def\cupress{Cambridge University Press}
\def\pup{Princeton University Press}
\def\wss{World Scientific, Singapore}
\def\oup{Oxford University Press}
\def\asj{Astron.~J}
\def\imp{Int.\ J.\ Mod.\ Phys.}


\begin{thebibliography}{99}
\bibitem{Sunyaev:72}
\jnl{R.A.~Sunyaev and Ya.~Zel'dovich}{1972}{}{Comm.\ Astrophys. \
Space Phys.}{4}{173}
\bibitem{Sunyaev:80}
\jnl{R.A.~Sunyaev and Ya.~Zel'dovich}{1980}{}{\mn}{190}{143}
\bibitem{Birkinshaw:86}
\jnl{M.~Birkinshaw and A.T.~Moffet}{1986}{}{\em
Highl.~Astron.}{7}{321}
\bibitem{Birkinshaw:91}
\proc{M.~Birkinshaw}{}{Physical Cosmology}{1991}{A. Blanchard
\et}{Editions Frontieres, Gif sur Yvette, France, 1991}
\bibitem{Hughes:98}
\jnl{J.P.~Hughes and M.~Birkinshaw}{1998}{}{\apj}{497}{645}
\bibitem{Jones:93}
\jnl{M.E.~Jones \et}{1993}{}{\it Nature}{365}{320}
\bibitem{Grainge:93}
\jnl{K.~Grainge \et}{1993}{}{\mn}{265}{L57}
\bibitem{Carlstrom:96}
\jnl{J.~Carlstrom, M.~Joy, and L.E.~Grego}{1996}{}{\apjl}{456}{L75}
\bibitem{Grainge:96}
\jnl{K.~Grainge \et}{1996}{}{\mn}{278}{L17}
\bibitem{Carlstrom:98}
\proc{J.E.~Carlstrom \et}{}{Eighteenth Texas Symposium on
Relativistic Astrophysics and Cosmology}{1998}{A.~Olinto, J.~Frieman, and
D.~Schramm}{{\wss}, 1998}
\bibitem{Reese:00}
\jnl{E.D.~Reese \et}{2000}{}{\apj}{533}{38}
\bibitem{Grego:00}
\jnl{L.~Grego \et}{2000}{}{\apj}{539}{39}
\bibitem{Patel:00}
\jnl{S.K. Patel \et}{2000}{}{\apj}{541}{37}
\bibitem{Joy:01}
\jnl{M.~Joy \et}{2001}{}{\apjl}{551}{L1}
\bibitem{Grego:01}
\jnl{L.~Grego \et}{2001}{}{\apj}{552}{2}
\bibitem{Bolocam:1}
\proc{P.D.~Mauskopf \et}{}{Imaging at Radio through Submilimeter
Wavelength}{1999}{J.~Mangum}{The Astronomical Society of the Pacific, Salt
Lake City, 2000}
\bibitem{Bolocam:2}
\proc{J.~Glenn \et}{}{Advanced Technology MMW, Radio, and Terahertz
Telescopes}{1998}{T.G.~Phillips, SPIE
Proc.~Vol.~3357}{SPIE-International Society for Optical
Engeneering. Bellingham, WA, 1998}
\bibitem{AMI:1}
\jnl{R.~Kneissl {\it et al.}}{2001}{}{\mn}{328}{783}
\bibitem{AMI:2}
\proc{M.E.~Jones}{}{AMiBA 2001: High-Z Clusters, Missing Baryons, and
CMB Polarization}{2001}{L.~W.~Chen}{Astronomical Society of the
Pacific, Salt Lake City 2002}
\bibitem{SZA:1}
\jnl{G.P.~Holder {\it et al.}}{2000}{}{\apj}{544}{629}
\bibitem{SZA:2}
\proc{J.J.~Mohr {\it et al.}}{}{Extrasolar Planets in Cosmology: The
VLT Opening Symposium}{2000}{A.~Renzini}{Springer, Berlin, 2000}
\bibitem{SZA:3}
\prep{J.E.~Carlstrom {\it et al.}}{}{}{astro-ph/0103480}
\bibitem{Amiba}
\proc{K.~Y.~Lo {\it et al.}}{}{`New Cosmological Data and the Values
of the Fundamental Parameters'}{2000}{A.~Lasenby and
A.~Wilkinson}{Astronomical Society of the Pacific, Salt Lake City,
2002}
\bibitem{OCRA}
\proc{I.W.A.~Browne {\it et al.}}{}{'Radio
Telescopes'}{2000}{H.R.~Butcher, SPIE
Proc. Vol.~4015}{SPIE-International Society for Optical Engineering,
Bellingham, WA, 2000} 
\bibitem{SPT:1}
\jnl{S.~Majumdar and J.J.~Mohr}{2003}{}{\apj}{585}{603}
\bibitem{SPT:2}
\prep{J.J.~Mohr \et}{}{}{astro-ph/0208102}
\bibitem{ACT}
See at: http://www.hep.upenn.edu/${}^\sim$angelica/act/act.html
\bibitem{APEX}
See at: http://www.mpifr-bonn.mpg.de/div/mm/apex.html
\bibitem{VSA}
\prep{R.A. Watson \et}{}{}{astro-ph/0205378} 
\bibitem{CBI}
\jnl{B.S.~Mason \et}{2003}{}{\apj}{591}{540}
\bibitem{Aghanim:97}
\jnl{N.~Aghanim \et}{1997}{}{\aap}{325}{9}
\bibitem{Kay:01}
\jnl{S.T.~Kay, A.R.~Liddle and P.A.~Thomas}{2001}{}{\mn}{325}{835}
\bibitem{BWproc}
\prep{R.A.~Battye and J.~Weller}{}{}{astro-ph/0305465}
\bibitem{Thomas:89}
\jnl{P.~Thomas and R.G.~Carlberg}{1989}{}{\mn}{240}{1009}
\bibitem{Oukbir:92}
\jnl{J.~Oukbir and A.~Blanchard}{1992}{}{\aap}{262}{L210}
\bibitem{Scaramella:93}
\jnl{R.~Scaramella, R.~Cen, and J.P.~Ostriker}{1993}{}{\apj}{416}{399}
\bibitem{Barbosa:96}
\jnl{D.~Barbosa \et}{1996}{}{\aap}{314}{13}
\bibitem{Viana:96}
\jnl{P.T.P.~Viana and A.R.~Liddle}{1996}{}{\mn}{281}{323}
\bibitem{Eke:96}
\jnl{V. Eke, S. Cole and C. Frenk}{1996}{}{\mn}{282}{263}
\bibitem{Wang:98}
\jnl{L.~Wang and P.~J.~Steinhardt}{1998}{}{\apj}{508}{483}
\bibitem{Pierpaoli:01}
\jnl{E.~Pierpaoli, D.~Scott and M.~White}{2001}{}{\mn}{325}{77}
\bibitem{Haiman:01a}
\jnl{Z.~Haiman, J.J.~ Mohr and G.P.~Holder}{2001}{}{\apj}{553}{545}
\bibitem{Holder:01}
\jnl{G.~Holder, Z.~Haiman, and J.J.~Mohr}{2001}{}{\apjl}{560}{L111}
\bibitem{Benson:02}
\jnl{A.J.~Benson, C.~Reichardt, and
M.~Kamionkowski}{2002}{}{\mn}{331}{71}
\bibitem{Weller:02b}
\jnl{J.~Weller, R.A.~Battye, and R.~Kneissl}{2002}{}{\prl}{88}{231301}
\bibitem{Perlmutter:97}
\jnl{S.~Perlmutter \et}{1997}{}{\apj}{483}{565}
\bibitem{Riess:98}
\jnl{A.~Riess \et}{1998}{}{\asj}{116}{1009}
\bibitem{Perlmutter:99a}
\jnl{S.~Perlmutter \et}{1999}{}{\apj}{517}{565}
\bibitem{Riess:01}
\jnl{A.~Riess \et}{2001}{}{\apj}{560}{49}
\bibitem{Wetterich:88}
\jnl{C.~Wetterich}{1988}{}{\np}{B302}{668}
\bibitem{Ratra:88}
\jnl{P.J.E.~Peebles and B.~Ratra}{1988}{}{\apj}{325}{L17}
\bibitem{Peebles:88}
\jnl{B.~Ratra and P.~J.~E.~Peebles}{1988}{}{\pr}{D37}{3406}
\bibitem{Wetterich:95}
\jnl{C.~Wetterich}{1995}{}{\aap}{301}{321}
\bibitem{Frieman:95}
\jnl{J.~Frieman \et}{1995}{}{\prl}{75}{2077}
\bibitem{Coble:97}
\jnl{K.~Coble, S.~Dodelson and J.A.~Frieman}{1997}{}{\pr}{D55}{1851}
\bibitem{Ferreira:97}
\jnl{P.G.~Ferreira and M.~Joyce}{1997}{}{\prl}{79}{4740}
\bibitem{Copeland:98}
\jnl{E.J.~Copeland, A.R.~Liddle and D.~Wands}{1998}{}{\pr}{D57}{4686}
\bibitem{Caldwell:98}
\jnl{R.R.~Caldwell, R.~Dave and
P.J.~Steinhardt}{1998}{}{\prl}{80}{1582}
\bibitem{Zlatev:99}
\jnl{I.~Zlatev, L.~Wang and P.J.~Steinhardt}{1999}{}{\prl}{82}{896}
\bibitem{Steinhardt:99}
\jnl{P.J.~Steinhardt, L.~Wang and I.~Zlatev}{1999}{}{\pr}{D59}{123504}
\bibitem{Brax:99}
\jnl{P.~Brax and J.~Martin}{1999}{}{\pl}{B 468}{40}
\bibitem{AS:00}
\jnl{A.~Albrecht and C.~Skordis}{2000}{}{\prl}{84}{2076}
\bibitem{Efstathiou:99}
\jnl{G.~Efstathiou}{1999}{}{\mn}{310}{842}
\bibitem{Huterer:99}
\jnl{D.~Huterer and M.S.~Turner}{1999}{}{\pr}{D 60}{081301}
\bibitem{Saini:99}
\jnl{T.D.~Saini \et}{2000}{}{\prl}{85}{1162}
\bibitem{Maor:01}
\jnlerr{I.~Maor, R.~Brustein and
P.J.~Steinhardt}{2001}{}{\prl}{86}{6}{87}{0499901}
\bibitem{Astier:00}
\prep{P.~Astier}{}{}{astro-ph/0008306}
\bibitem{Weller:01}
\jnl{J.~Weller and A.~Albrecht}{2001}{}{\prl}{86}{1939}
\bibitem{Weller:01a}
\jnl{J.~Weller and A.~Albrecht}{2001}{}{\pr}{D65}{103512}
\bibitem{Sahni:02}
\jnl{V.~Sahni \et}{2003}{}{JETP Lett.}{77}{201}
\bibitem{Huterer:02}
\jnl{D.~Huterer and G.~Starkman}{2003}{}{\prl}{90}{031301}
\bibitem{class1}
\jnl{K.-H. Chae {\it et al}}{2002}{}{\prl}{89}{151301}
\bibitem{class2}
\prep{K.-H. Chae}{2002}{}{}{astro-ph/0211244}
\bibitem{SDSS:1}
\proc{J.E.~Gunn and G.R.~Knapp}{}{Sky Surveys: Protostars to
Protogalaxies}{1993}{B.T.~Soifer, ASP Conf.~Ser.~43}{ASP. San
Francisco, CA ,1993}
\bibitem{SDSS:2}
\proc{J.E.~Gunn and D.H.~Weinberg}{}{Wide Field Spectroscopy and the
Distant Universe}{1995}{S.J.~Maddox and A.~Aragon-Salamanca}{World
Scientific. Singapore, 1995} 

\bibitem{VISTA}
\proc{S.P.~Worsick \et}{}{Optical Design, Materials, Fabrication, and
Maintenance}{2000}{P.~Dierickx, SPIE Proc.~Vol.~4003}{SPIE-International Society for Optical
Engeneering. Bellingham, WA, 2000}




\bibitem{Evrard:02}
\jnl{A.E.~Evrard \et}{2002}{}{\apj}{573}{7}

\bibitem{Seljak:96}
\jnl{U.~Seljak and M.~Zaldarriaga}{1996}{}{\apj}{469}{437}

\bibitem{Copi:95}
\jnl{C.~Copi, D.N.~Schramm and M.S.~Turner}{1995}{}{Science}{267}{192}
\bibitem{Burles:98}
\jnl{S.~Burles and D.~Tytler}{1998}{}{\apj}{507}{732}

\bibitem{wmap}
\prep{G.~Hinshaw \et}{}{}{astro-ph/0302217}
\bibitem{Seljak:01}
\prep{U.~Seljak}{}{}{astro-ph/0111362}
\bibitem{Viana:02}
\jnl{P.T.P.~Viana, R.C.~Nichol and
A.R.~Liddle}{2002}{}{\apjl}{569}{L75}

\bibitem{Allen:02b}
\prep{S.W.~Allen \et}{}{}{astro-ph/0208394}

\bibitem{Freedman:01}
\jnl{W.L.~Freedman \et}{2001}{}{\apj}{553}{47}

\bibitem{Bean:02}
\jnl{R.~Bean and A.~Melchiorri}{2002}{}{\pr}{D65}{041302}
\bibitem{Lewis:02}
\jnl{A.~Lewis and S.~Bridle}{2002}{}{\pr}{D66}{103511}

\bibitem{Allen:02a}
\jnl{S.W.~Allen, R.W.~Schmidt and A.C.~Fabian}{2002}{}{\mn}{334}{L11}

\bibitem{Bridle:03}
S.~L.~Bridle, {\em private communication}

\bibitem{Garnavich:98}
\jnl{P.M.~Garnavich \et}{1998}{}{\apj}{509}{74}
\bibitem{Perlmutter:99b}
\jnl{S.~Perlmutter, M.~S.~Turner and M.~White}{1999}{}{\prl}{83}{670}
\bibitem{Press:74}
\jnl{W.H.~Press and P.~Schechter}{1974}{}{\apj}{187}{452}
\bibitem{Sheth:01}
\jnl{R.K.~Sheth and G.~Tormen}{2001}{}{\mn}{323}{1}
\bibitem{Jenkins:01}
\jnl{A.~Jenkins \et}{2001}{}{\mn}{321}{372}

\bibitem{White:01}
\jnl{M.~White}{2001}{}{\aap}{367}{27}

\bibitem{White:02}
\jnl{M.~White}{2002}{}{\apj Supp.}{143}{241}

\bibitem{Birkinshaw:99}
\jnl{M.~Birkinshaw}{1999}{}{\prp}{310}{97}

\bibitem{Battye:02}
R.A.~Battye and J.~Weller, {\em In preparation}

\bibitem{Mohr:99}
\jnl{J.J.~Mohr, B.~Mathiesen and A.E.~Evrard}{1999}{}{\apj}{517}{627}

\bibitem{Verde:01}
\jnl{L.~Verde, Z.~Haiman and D.~N.~Spergel}{2002}{}{\apj}{581}{5}

\bibitem{Lahav:91}
\jnl{O.~Lahav {\it et al.}}{1991}{}{\mn}{251}{128}

\bibitem{Lilje:92}
\jnl{P.B.~Lilje}{1992}{}{\apjl}{386}{L33}
\bibitem{Bryan:98}
\jnl{G.L.~Bryan and M.L.~Norman}{1998}{}{\apj}{495}{80}
\bibitem{Afshordi:01}
\jnl{N.~Afshordi and R.~Cen}{2002}{}{\apj}{564}{669}

\bibitem{Patridge:67}
\jnl{R.~B.~Patridge and P.~J.~E.~Peebles}{1967}{}{\apj}{147}{868}
\bibitem{Gunn:72}
\jnl{J.~E.~Gunn and J.~R.~Gott}{1972}{}{\apj}{176}{1}

\bibitem{Navarro:97}
\jnl{J.~F.~Navarro, C.~S.~Frenk and S.~D.~White}{1997}{}{\apj}{490}{493}

\bibitem{Bartlett:00}
\prep{J.G.~Bartlett}{}{}{astro-ph/0001267}
\bibitem{Cavaliere:76}
\jnl{A.~Cavaliere and R.~Fusco-Femiano}{1976}{}{\aap}{49}{137}
\bibitem{Molnar:00}
\jnl{S.M.~Molnar and M.~Birkinshaw}{2000}{}{\apj}{537}{542}

\bibitem{SPT:3}
J.E.~Carlstrom, {\em private communication}.

\bibitem{Cash:79}
\jnl{W.~Cash}{1979}{}{\apj}{228}{939}

\bibitem{Christensen:00}
\prep{N.~Christensen and R.~Meyer}{}{}{astro-ph/0006401}
\bibitem{Christensen:01}
\jnl{N.~Christensen \et}{2001}{}{\cqg}{18}{2677}

\bibitem{Sugiyama:95}
\jnl{N.~Sugiyama}{1995}{}{\apj~S.}{100}{281}
\bibitem{Viana:99}
\jnl{P.T.P.~Viana and A.R.~Liddle}{1999}{}{\mn}{303}{535}

\bibitem{SNAP}
\proc{J. Aldering \et}{}{Future Research Direction and Visions for Astronomy}{2002}{A.M.~Dressler, SPIE Proc.~Vol.~4835}{SPIE-International Society for Optical Engeneering. Bellingham, WA, 2002}

\bibitem{Mohr:97}
\jnl{J.J.~Mohr and A.E.~Evrard}{1997}{}{\apj}{517}{627}
\bibitem{Muanwong:01}
\prep{O.~Muanwong \et}{}{}{astro-ph/0102048}
\bibitem{Xu:01}
\prep{H.~Xu, G.~Jin and X.~Wu}{}{}{astro-ph/0101564}
\bibitem{Finoguenov:01}
\jnl{A.~Finoguenov, T.H.~Reiprich and H.~B\"oringer}{2001}{}{\aap}{369}{749}

\bibitem{Pierpaoli:02}
\jnl{E.~Pierpaoli \et}{2003}{}{\mn}{342}{163}
\bibitem{Huterer:02a}
\jnl{D.~Huterer and M.~White}{2002}{}{\apjl}{95}{578}
\bibitem{CBI:SZ}
\prep{J.R.~Bond \et}{}{}{astro-ph/0205386}
\bibitem{Komatsu:02}
\prep{E.~Komatsu and U.~Seljak}{}{}{astro-ph/0205468}


\bibitem{Hu:03}
\jnl{W.~Hu}{2003}{}{\pr}{D 67}{081304}
\end{thebibliography}
\end{document}